\documentclass[submit]{smj}

\usepackage{subfigure}
\usepackage{graphicx}
\usepackage{amssymb}
\usepackage{amsmath}
\usepackage{booktabs}
\usepackage{footnote}
\usepackage{appendix}

\usepackage{epstopdf}
\usepackage{subfigure}
\usepackage{wasysym}
\usepackage{graphicx,color}
\usepackage{multirow,fixltx2e}

\newtheorem{thm}{Theorem}[section]
\newtheorem{cor}{Corollary}[thm]

\Author{Fernanda V. Paula\Affil{1}, 
        Abra\~ao D. C. Nascimento\Affil{2} 
        and Get\'ulio J. A. Amaral\Affil{2}
}
\AuthorRunning{Fernanda V. Paula \textrm{et al.}}

\Affiliations{

\item Colegiado de Matem\'atica, 
      Universidade Federal do Tocantins, 
      Aragua\'ina, 
      TO, Brazil

\item Departamento de Estat\'istica,
      Universidade Federal de Pernambuco, 
      Recife,
      PE, Brazil

}   

\CorrAddress{Fernanda Vital de Paula, 
             Colegiado de Matem\'atica, 
             Universidade Federal do Tocantins, 
             Aragua\'ina, TO, 
             $77838-824$, Brazil}
\CorrEmail{fernandavital@uft.edu.br}
\CorrPhone{(+55)\;63\;2112\;2227}
\CorrFax{(+55)\;63\;2112\;2349}

\Title{A new extended Cardioid model: an application to wind data}
\TitleRunning{A new extended Cardioid model: an application to wind data}

\Abstract{
The Cardioid distribution is a relevant model for circular data. However, this model is not suitable for scenarios were there is asymmetry or multimodality. In order to overcome this gap, an extended Cardioid model is proposed, which is called Exponentiated Cardioid (EC) distribution. Besides, some of its properties are derived, such as trigonometric moments, kurtosis and skewness. A discussion about the modality and and expressions for the quantiles through approximations of the studied model are also presented. To fit the EC model, two estimation methods are presented based on maximum likelihood and quantile least squares procedures. The performance of proposed estimators is evaluated in a Monte Carlo simulation study, adopting both bias and mean square error as comparison criteria. Finally, 
the proposed model is applied to a dataset in the wind direction context.  Results indicate that the EC distribution may outperform Cardioid and the von Mises distributions.  
}


\begin{document}

\maketitle

\section{Introduction}

Circular data have been obtained from various fields, where the measurements are angle and directions, such as in Biology [\cite{Bat}], Zoology [\cite{Zoo}], Geology [\cite{Geo}] and others. Some examples are related to birds navigational, variation in the onset of leukaemia, orientation data in textures and wind directions. The periodic nature of circular data imposes a specific treatment which is appropriate for non-euclidean space. Even though symmetry is assumed by several circular models, there are many practical situations where asymmetric distributions are necessary. Thus, a new tractable flexible circular model will be addressed in this paper. 


There are some generators for representing circular and directional data. 
Among them, the simplest is known as \emph{perturbation procedure} proposed by \cite{Jeffreys} and it is based on the product of an existing circular density and a function chosen such that the resulting expression is also a circular density [\cite{Pewseybook}]. The cardioid (C) and sine-skewed distributions [\cite{Sine}] are particular cases of this method. The C distribution was introduced by \cite{Jeffreys} as cosine perturbation of the continuous circular uniform distribution and has probability density function (pdf) given by
\[
f_C(\theta)=\frac{1}{2\pi}\{1+2 \rho \cos(\theta-\mu)\},
\]
where $0 \leq \mu < 2 \pi$, $|\rho| \leq \frac{1}{2}$ and $\rho$ and $\mu$ are concentration and mean direction parameters, respectively. Further, the circular uniform distribution is obtained when $\rho=0$. Other generator is stemmed from the real line around the circumference, called \emph{wrapping models} [\cite{Jama}]. The wrapped Cauchy and normal models are examples of this method. Generators defined by transforming the argument of some existing densities, say $g(\theta)$, replacing its argument, $\theta$, by functions of it, can also be mentioned. Some distribution generators in this way are \cite{Abe}, \cite{Jones}, \cite{Papa} (generated from C) and \cite{Bat} (from von Mises) families. Moreover, the von Mises distribution is particularly useful in this paper because its wide application to circular data. Its pdf is given by (for $0 \leq \theta < 2 \pi$)
\[
f_V(\theta)=[2\pi I_0(\rho)]^{-1}\exp[\rho\cos(\theta-\mu)],
\]
where $0 \leq \mu < 2 \pi$, $\rho \geq 0$ and $I_0(\rho)=\int_{0}^{2\pi}\exp[\rho\cos(\phi-\mu)]d\phi$ is the modified Bessel function of the first kind and order zero.

A further form to construct a circular model is the called \emph{M\"obius transformation}. An example is the proposed family in \cite{Kato}, derived from the von Mises distribution. Another method is based on the transformation of a bivariate linear random variable to its directional component. The obtained models are called  \emph{offset} distributions [\cite{Mardia}].

The previous generators result generally in symmetrical distributions. However, tractable circular models with asymmetric shape are required in several applications into the circular context. Some asymmetric extensions have been proposed in the literature. Among them, finite mixtures of unimodal models[ \cite{Mardia}] and the application of multiplicative mixing, as that used by \cite{Gatto} to extended the von Mises law. \cite{Pewseybook} also made reference to the non-negative trigonometric moment distributions [\cite{Duran}]. 


In Euclidean space, there are several ways to extend well-defined models. One of them is by exponentiating a cumulative distribution
function (cdf), say $F$, by a positive real number $\beta$, $F(\cdot)^\beta$ see; AL-Hussaini and Ahsanullah \cite{Exponentiated}. As far as we know, this approach has not been previously used in the circular data context.

In this paper, the methodology described by AL-Hussaini and Ahsanullah \cite{Exponentiated} is used to obtain a new circular distribution, called the \emph{Exponentiated Cardioid} model. The additional parameter may add events of amodality and bimodality to the baseline unimodal C, as will be shown. Expressions for its trigonometric moments are also proposed. Two estimation procedures for the EC parameters are presented: maximum likelihood estimator (MLE) and quantile least squares estimator (QLSE). In order to compare those estimators, a Monte Carlo simulation study is performed, on which we conclude that ML estimates have smaller mean square errors than those of QLS estimates in almost all considered cases. Finally, in order to illustrate the EC distribution potentiality, a comparison of its fit to those due to C and von Mises distributions is provided. Results of Kuiper and Watson goodness-of-fit statistics present evidence in favor of our proposal.



This paper is organized as follows. In Sections 2 and 3, the EC distribution and some of its mathematical properties are presented. Section 4 deals with estimation procedures. Finally, numerical results obtained from real data studies are presented

\section{The Proposed Model}
The EC ditribution has cdf given by (for $0<\theta\leq2\pi$)
\begin{equation}
F(\theta;\beta,\rho,\mu)=\left\{\frac{\theta}{2\pi}+\frac{\rho}{\pi}[\sin(\theta-\mu)+\sin (\mu)]\right\}^{\beta},
\label{acum}
\end{equation}
where $0<\mu\leq2\pi$, $0\leq\rho\leq0.5$ and $\beta>0$ and, therefore, its pdf is
\begin{equation}
f(\theta;\beta,\rho,\mu)=\left\{\frac{\theta}{2\pi}+\frac{\rho}{\pi}[\sin(\theta-\mu)+\sin(\mu)]\right\}^{\beta-1}\frac{\beta}{2\pi}\{[1+2 \rho \cos(\theta-\mu)]\}.
\label{density}
\end{equation}
This situation is denoted as $\Theta\sim EC(\beta,\rho,\mu)$. As special cases, the C distribution follows for $\beta=1$ and the uniform distribution is obtained when $\beta=1$ and $\rho=0$. In this position, it is important to mention that the Cardioid extension by exponentiation requires to the replacement zero by $2\pi$ in the EC support comparatively to that of the C model. This change avoids an undefinition at zero.

Figure 1 displays EC pdf curves and their associated histograms drawn from generated data for several parametric points. It is noticeable bimodal and asymmetric events in contrast with its baseline at $\beta=1$. Moreover, higher values of $\rho$ indicate more concentrated scenarios, as illustrated in Figure 1(d).

\begin{figure}[htp]
	\centering
	\subfigure[For $\rho=0.2$ and $\mu=2$.]{
		\resizebox*{5cm}{5cm}{\includegraphics{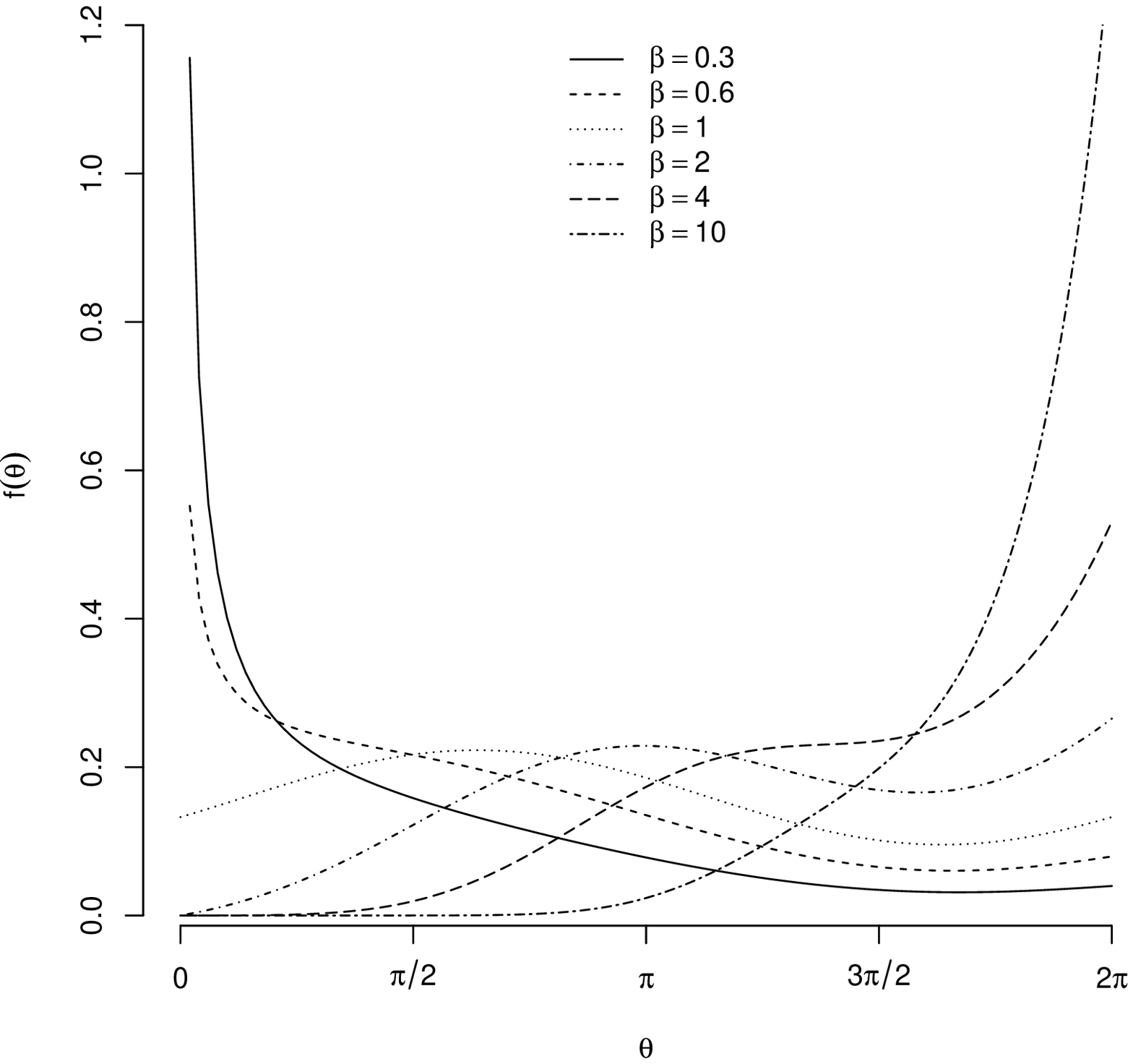}}}
	\centering
	\subfigure[For $\rho=0.2$ and $\mu=2$.]{
		\resizebox*{5cm}{5cm}{\includegraphics{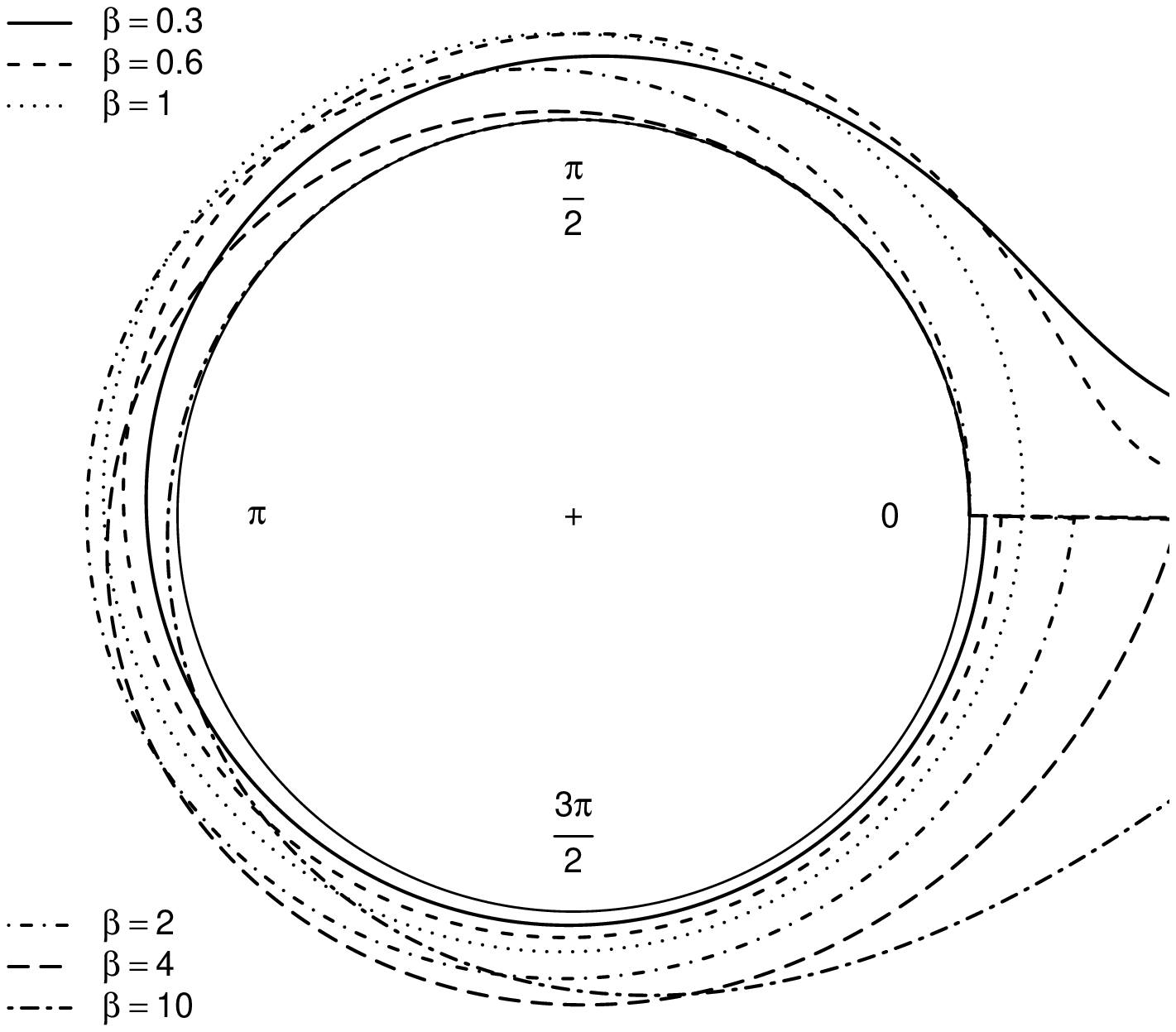}}}
	\centering
	\subfigure[For $\beta=2$ and $\mu=2$.]{
		\resizebox*{5cm}{5cm}{\includegraphics{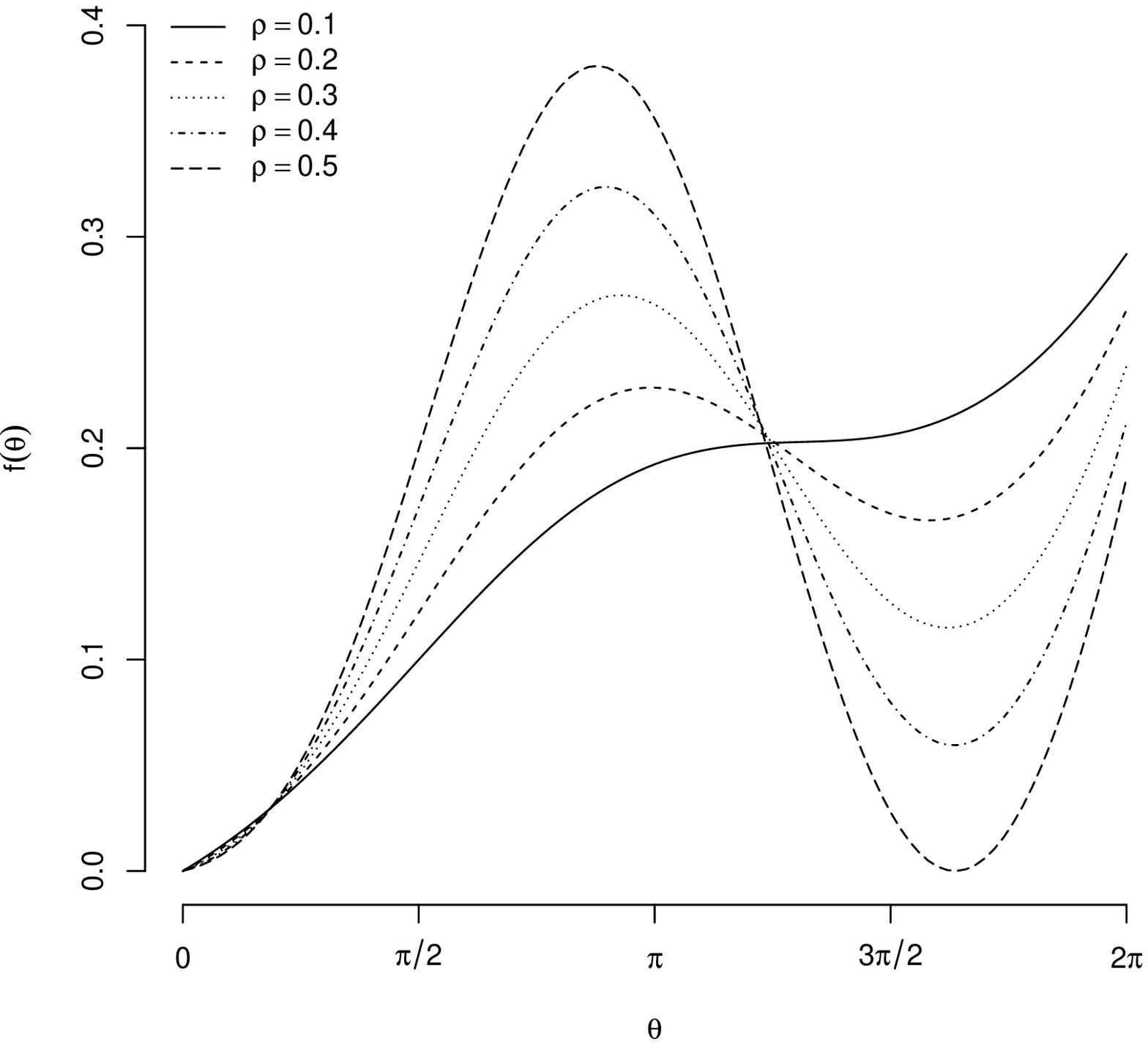}}}
	\subfigure[For $\beta=2$ and $\mu=2$.]{
		\resizebox*{5cm}{5cm}{\includegraphics{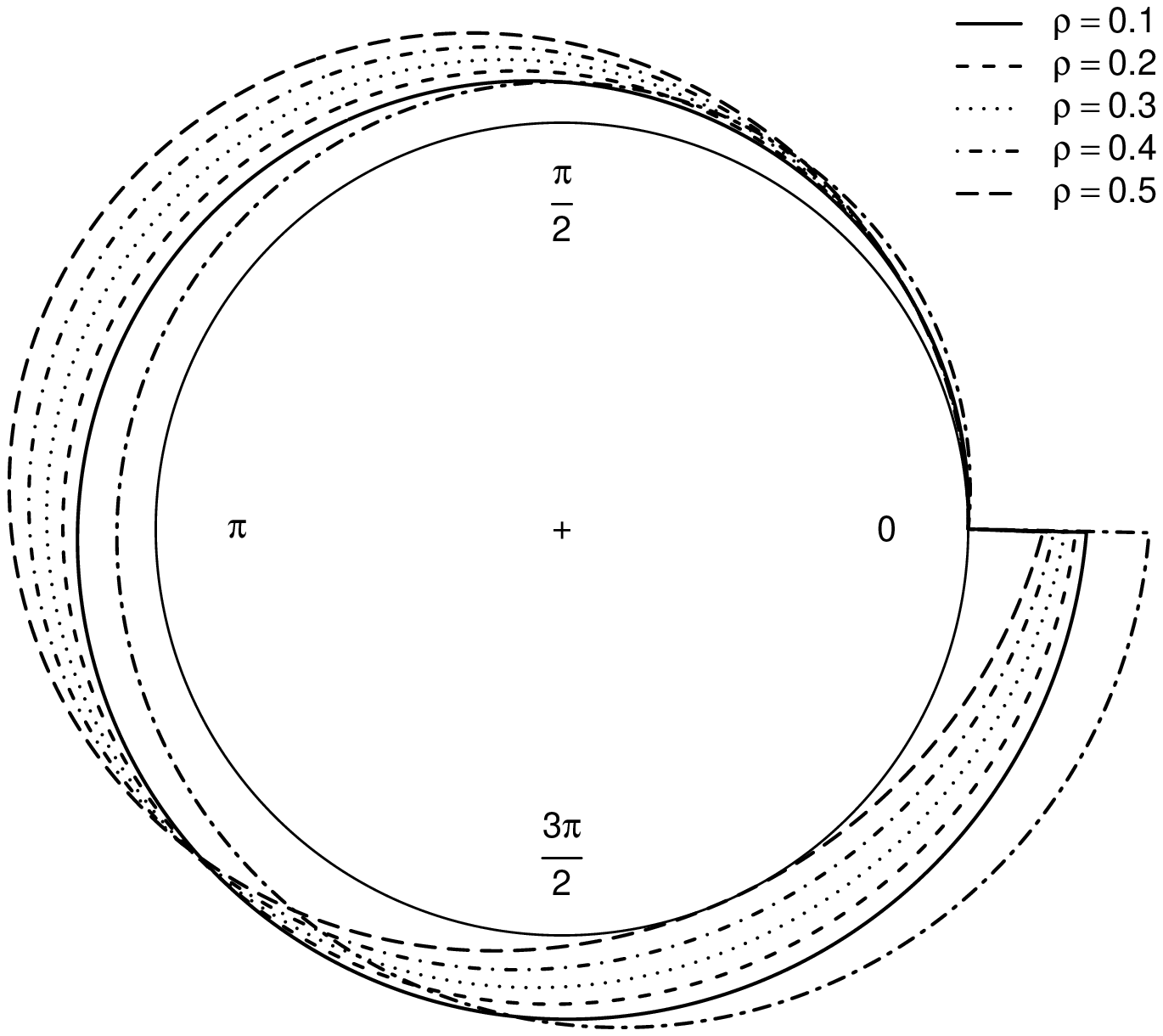}}}
	\centering
	\subfigure[For $\beta=2$ and $\rho=0.2$.]{
		\resizebox*{5cm}{5cm}{\includegraphics{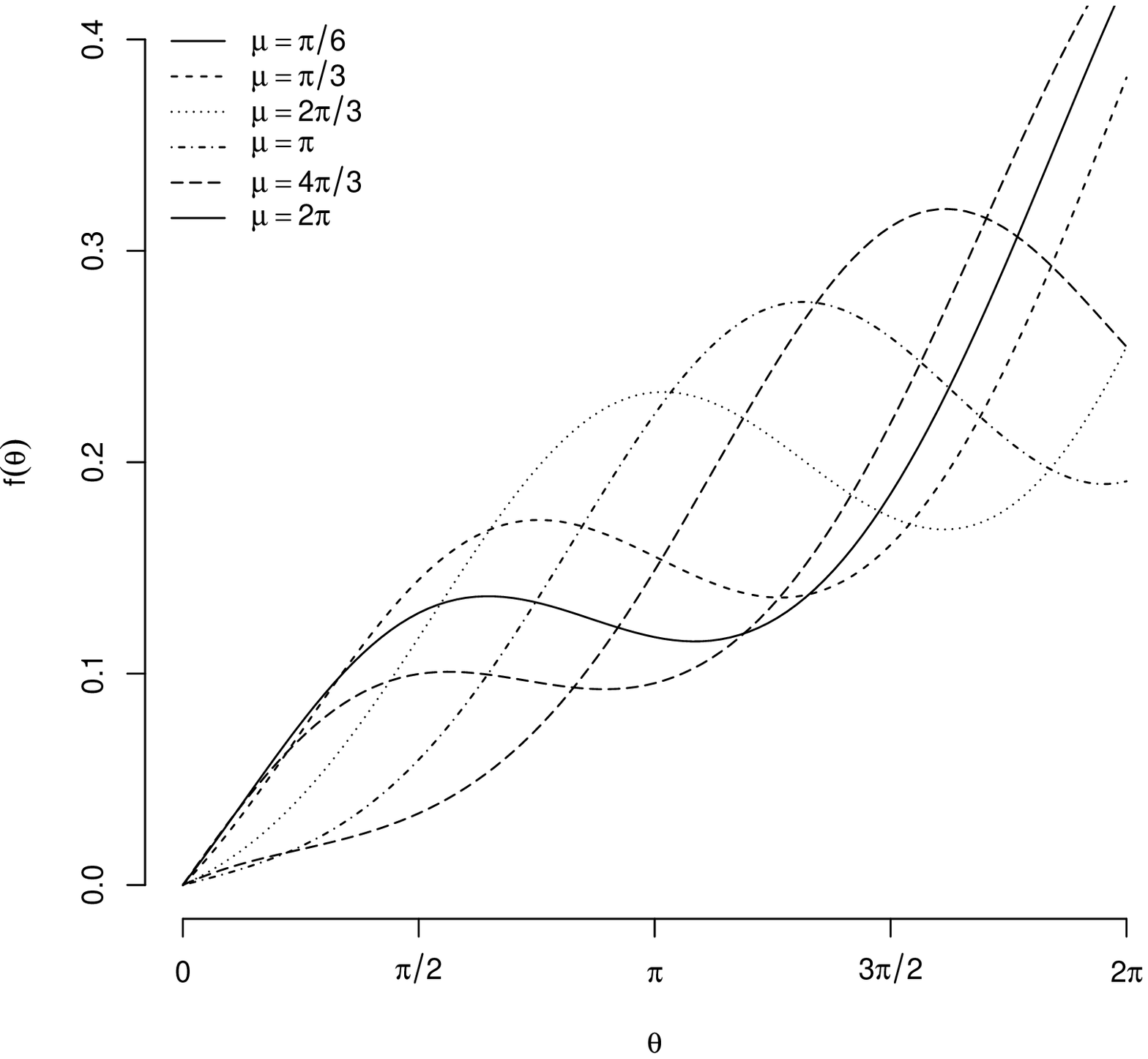}}}
	\subfigure[For $\beta=2$ and $\rho=0.2$.]{
		\resizebox*{5 cm}{5cm}{\includegraphics{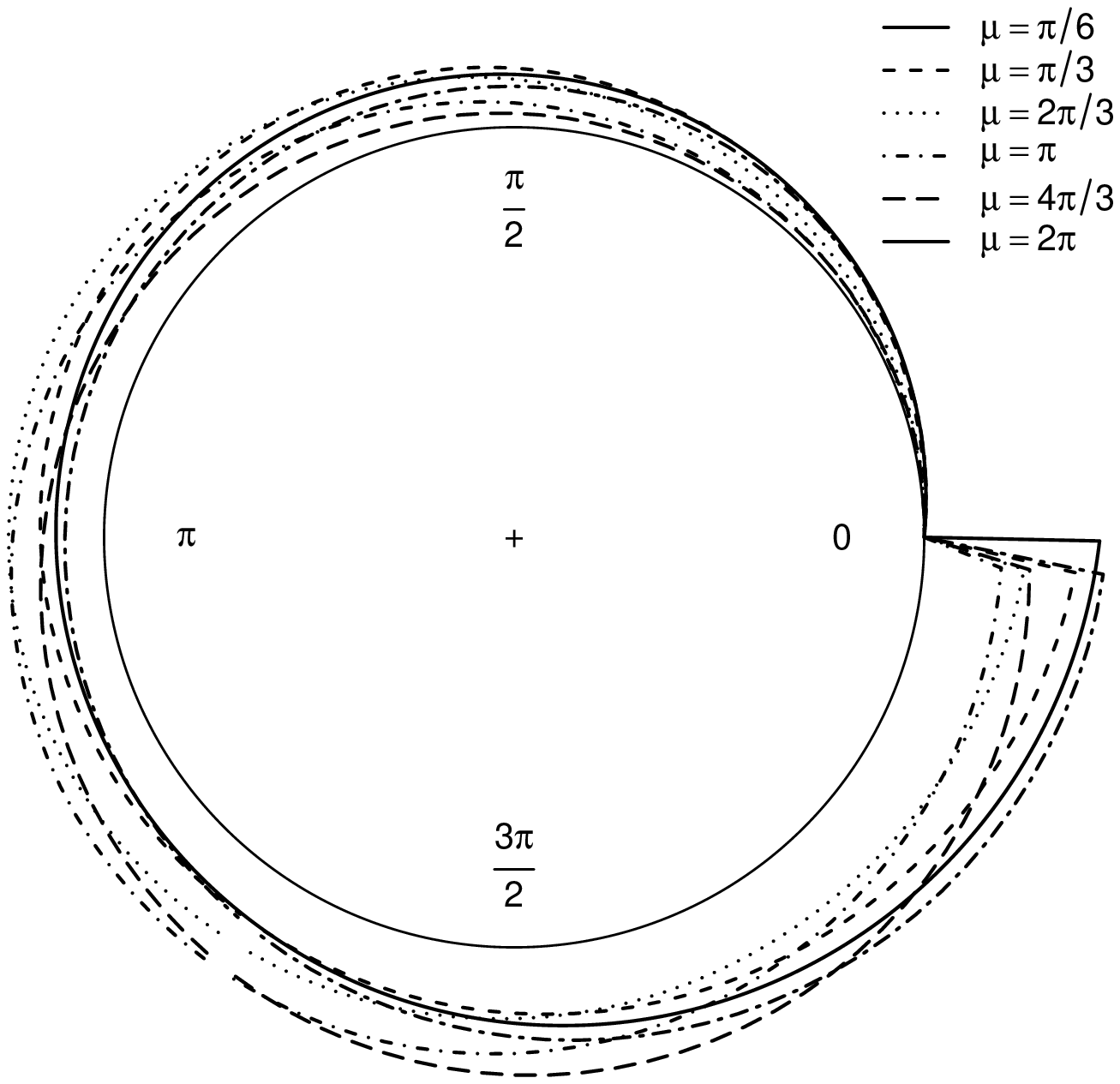}}}
	\caption{Theoretical and empirical EC densities for some parametric points.} 
	\label{plot}
\end{figure}

\subsection{Quantiles}


\begin{thm}
	Let $\Theta\sim EC(\beta,\rho,\mu)$. The following expressions are valid for quantiles approximation:
	\begin{enumerate}
		\item For $Q_{\alpha}-\mu \in [0,0.60]$:
		\[
		Q_{\alpha}\approx\frac{2\pi}{1+2\rho}\left\{\alpha^{\frac{1}{\beta}}-\frac{\rho}{\pi}\left[\sin(\mu)-\mu\right]\right\};
		\]
		\item For $Q_{\alpha}-\mu \in (0.60,2.62]$ and $\rho\neq0$:
		\[
		Q_{\alpha}\approx\frac{\pi}{\rho}\left[D-\sqrt{E}\right];
		\]
		\item For $Q_{\alpha}-\mu \in (2.62,3.64]$:
		\[
		Q_{\alpha}\approx\frac{2\pi}{1-2\rho}\left\{\alpha^{\frac{1}{\beta}}-\frac{\rho}{\pi}\left[\sin(\mu)+\mu+\pi\right]\right\};
		\]
		\item For $Q_{\alpha}-\mu \in (3.64,5.76]$ and $\rho\neq0$:
		\[
		Q_{\alpha}\approx-\frac{\pi}{\rho}\left[G-\sqrt{H}\right];
		\]
		\item For $Q_{\alpha}-\mu \in (5.76,6.28]$:
		\[
		Q_{\alpha}\approx\frac{2\pi}{1+2\rho}\left\{\alpha^{\frac{1}{\beta}}-\frac{\rho}{\pi}\left[\sin(\mu)-\mu-2\pi\right]\right\};
		\]
	\end{enumerate}
	where
	$C=\alpha^\frac{1}{\beta}-\frac{\rho}{\pi}\left\{\sin(\mu)-\frac{\mu}{2}[\pi+\mu]+\frac{8-\pi^2}{8}\right\}$,$D=\frac{1}{\pi}\left[\frac{1+\pi\rho+2\rho\mu}{2}\right]$, $E=D^2-\frac{2\rho C}{\pi}$, $G=D-2\rho\left[1+\frac{\mu}{\pi}\right]$, $H=G^2+\frac{2\rho F}{\pi}$ and $F=C-\rho\left[\frac{\mu^2-2}{\pi}+\frac{8\mu+5\pi^2}{4}\right]$.
	\label{quantiles}
\end{thm}

The EC quantile function (qf) is analytically intractable. However, from trigonometric results and the Taylor serie expansions, approximated expressions for the EC qf may be given, according to Theorem \ref{quantiles}. The proof is presented in Appendix \ref{quantilesproof}.



\subsection{Modality Essays}

The flexibility of the EC distribution is partially portrayed in Table \ref{modality}. It knows that the C model ($\beta=1$) is unimodal. In contrast, the EC distribution can be classified as amodal, unimodal and bimodal for different values of $\beta$, $\rho$ and $\mu$. This fact shows our proposal has greater flexibility than its corresponding baseline.

\begin{table}[htp]
\scriptsize 
\centering
\caption{Modality of the EC distribution, for different values ​​of the parameters, where $\varnothing$=Amodal, $\Diamond$=Unimodal e $\blacksquare$=Bimodal.}
{$\begin{tabular}{c|c}
		
\parbox[c][5.5cm][c]{6.5cm}{ $$\begin{tabular}{c|c|c|c|c|c} \hline
$\beta=0.3$ & \multicolumn{5}{c}{$\rho$}    
			
\\\hline 
$\mu$&  $\frac{1}{10}$ & $\frac{1}{5}$ & $\frac{3}{10}$ & $\frac{2}{5}$ & $\frac{1}{2}$ \\\hline
$\frac{\pi}{6}$ & $\Diamond$ & $\Diamond$ &  $\Diamond$ & $\Diamond$ & $\Diamond$\\\hline
$\frac{\pi}{3}$ & $\Diamond$ & $\Diamond$ &  $\Diamond$ & $\Diamond$ & $\Diamond$\\\hline
$\frac{2\pi}{3}$ & $\Diamond$ & $\Diamond$ & $\Diamond$ & $\Diamond$ & $\Diamond$\\\hline
$\pi$ & $\varnothing$ & $\varnothing$ & $\varnothing$ & $\varnothing$ & $\varnothing$\\\hline
$\frac{4\pi}{3}$ & $\varnothing$ & $\Diamond$ &  $\Diamond$ & $\Diamond$ & $\Diamond$\\\hline
$2\pi$ & $\Diamond$ & $\Diamond$ &  $\Diamond$ & $\Diamond$ & $\Diamond$\\\hline
\end{tabular}$$} &
		
\parbox[c][5.5cm][c]{6.5cm}{
$$\begin{tabular}{c|c|c|c|c|c}\hline
$\beta=0.6$& \multicolumn{5}{c}{$\rho$}\\ \hline 
{$\mu$} & $\frac{1}{10}$ & $\frac{1}{5}$ & $\frac{3}{10}$ & $\frac{2}{5}$ & $\frac{1}{2}$ \\\hline
$\frac{\pi}{6}$ & $\Diamond$ & $\Diamond$ &  $\Diamond$ & $\Diamond$ & $\Diamond$\\\hline
$\frac{\pi}{3}$ & $\Diamond$ & $\Diamond$ &  $\Diamond$ & $\Diamond$ & $\Diamond$\\\hline
$\frac{2\pi}{3}$ & $\Diamond$ & $\Diamond$ &  $\Diamond$ & $\Diamond$ & $\blacksquare$\\ \hline
$\pi$ & $\varnothing$ & $\Diamond$ & $\Diamond$ & $\Diamond$ & $\Diamond$\\\hline
$\frac{4\pi}{3}$ & $\Diamond$ & $\Diamond$ &  $\Diamond$ & $\Diamond$ & $\Diamond$\\\hline
$2\pi$ & $\Diamond$ & $\Diamond$ &  $\Diamond$ & $\Diamond$ & $\Diamond$\\\hline
\end{tabular}$$} \\ \hline
		
\parbox[c][5.5cm][c]{6.5cm}{
$$\begin{tabular}{c|c|c|c|c|c}\hline
$\beta=1.0$& \multicolumn{5}{c}{$\rho$}\\ \hline
$\mu$ & $\frac{1}{10}$ & $\frac{1}{5}$ & $\frac{3}{10}$ & $\frac{2}{5}$ & $\frac{1}{2}$ \\\hline
$\frac{\pi}{6}$ & $\Diamond$ & $\Diamond$ &  $\Diamond$ & $\Diamond$ & $\Diamond$\\\hline
$\frac{\pi}{3}$ & $\Diamond$ & $\Diamond$ &  $\Diamond$ & $\Diamond$ & $\Diamond$\\\hline
$\frac{2\pi}{3}$ & $\Diamond$ & $\Diamond$ &  $\Diamond$ & $\Diamond$ & $\Diamond$\\\hline
$\pi$ & $\Diamond$ & $\Diamond$ & $\Diamond$ & $\Diamond$ & $\Diamond$\\\hline
$\frac{4\pi}{3}$ & $\Diamond$ & $\Diamond$ &  $\Diamond$ & $\Diamond$ & $\Diamond$\\\hline
$2\pi$ & $\Diamond$ & $\Diamond$ &  $\Diamond$ & $\Diamond$ & $\Diamond$\\\hline
\end{tabular}$$} & 
		
\parbox[c][5.5cm][c]{6.5cm}{
$$\begin{tabular}{c|c|c|c|c|c}\hline
$\beta=2.0$& \multicolumn{5}{c}{$\rho$}\\ \hline
$\mu$ & $\frac{1}{10}$ & $\frac{1}{5}$ & $\frac{3}{10}$ & $\frac{2}{5}$ & $\frac{1}{2}$ \\\hline
$\frac{\pi}{6}$ & $\Diamond$ & $\blacksquare$ &  $\blacksquare$ & $\blacksquare$ & $\blacksquare$\\\hline
$\frac{\pi}{3}$ & $\Diamond$ & $\blacksquare$ &  $\blacksquare$ & $\blacksquare$ & $\blacksquare$\\\hline
$\frac{2\pi}{3}$ & $\Diamond$ & $\blacksquare$ &  $\blacksquare$ & $\blacksquare$ & $\blacksquare$\\\hline
$\pi$ & $\blacksquare$ & $\blacksquare$ &  $\blacksquare$ & $\blacksquare$ & $\Diamond$\\\hline
$\frac{4\pi}{3}$ & $\Diamond$ & $\Diamond$ &  $\Diamond$ & $\blacksquare$ & $\blacksquare$\\\hline
$2\pi$ & $\Diamond$ & $\blacksquare$ &  $\blacksquare$ & $\blacksquare$ & $\blacksquare$\\\hline
\end{tabular}$$} \\ \hline
		
\parbox[c][5.5cm][c]{6.5cm}{
$$\begin{tabular}{c|c|c|c|c|c}\hline
$\beta=4.0$& \multicolumn{5}{c}{$\rho$}\\ \hline
$\mu$ & $\frac{1}{10}$ & $\frac{1}{5}$ & $\frac{3}{10}$ & $\frac{2}{5}$ & $\frac{1}{2}$ \\\hline
$\frac{\pi}{6}$ & $\Diamond$ & $\Diamond$ &  $\blacksquare$ & $\blacksquare$ & $\blacksquare$\\\hline
$\frac{\pi}{3}$ & $\Diamond$ & $\Diamond$ &  $\blacksquare$ & $\blacksquare$ & $\blacksquare$\\\hline
$\frac{2\pi}{3}$ & $\Diamond$ & $\Diamond$ &  $\blacksquare$ & $\blacksquare$ & $\blacksquare$\\\hline
$\pi$ & $\Diamond$ & $\blacksquare$ &  $\blacksquare$ & $\blacksquare$ & $\Diamond$\\\hline
$\frac{4\pi}{3}$ & $\Diamond$ & $\Diamond$ &  $\Diamond$ & $\Diamond$ & $\Diamond$\\\hline
$2\pi$ & $\Diamond$ & $\Diamond$ &  $\blacksquare$ & $\blacksquare$ & $\blacksquare$\\\hline
\end{tabular}$$} &
		
\parbox[c][5.5cm][c]{6.5cm}{
$$\begin{tabular}{c|c|c|c|c|c}\hline
$\beta=10.0$& \multicolumn{5}{c}{$\rho$}\\ \hline
$\mu$ & $\frac{1}{10}$ & $\frac{1}{5}$ & $\frac{3}{10}$ & $\frac{2}{5}$ & $\frac{1}{2}$ \\\hline
$\frac{\pi}{6}$ & $\Diamond$ & $\Diamond$ &  $\Diamond$ & $\blacksquare$ & $\blacksquare$\\\hline
$\frac{\pi}{3}$ & $\Diamond$ & $\Diamond$ &  $\Diamond$ & $\blacksquare$ & $\blacksquare$\\\hline
$\frac{2\pi}{3}$ & $\Diamond$ & $\Diamond$ &  $\Diamond$ & $\blacksquare$ & $\blacksquare$\\\hline
$\pi$ & $\Diamond$ & $\Diamond$ &  $\Diamond$ & $\blacksquare$ & $\Diamond$\\\hline
$\frac{4\pi}{3}$ & $\Diamond$ & $\Diamond$ &  $\Diamond$ & $\Diamond$ & $\Diamond$\\\hline
$2\pi$ & $\Diamond$ & $\Diamond$ &  $\Diamond$ & $\blacksquare$ & $\blacksquare$\\\hline
\end{tabular}$$}
		
\end{tabular}$}
\label{modality}
\end{table}

\section{Moments}

Expressions for the first two trigonometric moments of the EC model are derived in this section. Moreover, standard descriptive measures for the proposed model are obtained from them.  In general, EC trigonometric moments do not present closed-form expressions. Thus, they are represented through expansions in terms of a proposed special function as follows.

\begin{thm}
	Let $\Theta\sim EC(\beta,\rho,\mu)$. Its cdf can be represented as
	\[
	F(\theta)=\sum_{k=0}^{+\infty}\sum_{s=0}^{k}T_{k,s}\theta^{\beta-k}[\sin(\theta-\mu)]^{s}\left\{[\sin(\theta-\mu)]^{k-2s}M_0+{[\sin(\mu)]^{k-2s}}M_1\right\},
	\]
	where $M_0=I(|\sin(\theta-\mu)|\geq|\sin(\mu)|)$, $M_1=I(|\sin(\theta-\mu)|<|\sin(\mu)|)$, $I(.)$ refers to the indicator function and $T_{k,s}$ is given by
	\[
	T_{k,s}(\beta,\rho,\mu)={\beta \choose k}{k \choose s}\left(\frac{1}{2\pi}\right)^{\beta-k}\left(\frac{\rho}{\pi}\right)^{k}[\sin(\mu)]^{s}.
	\]
	\label{theorem1}
\end{thm}

It is known, the $n$th central trigonometric moment of the EC is given by

\[
\mu_{p}=\mathbb{E}\{[\cos[p(\Theta-\mu)]\}+i\mathbb{E}\{[\sin[p(\Theta-\mu)]\}.
\]

Using integration by parts, it follows that

\begin{equation*}
\mathbb{E}\{[\cos[p(\Theta-\mu)]\}=\int_{0}^{2\pi} \cos[p(\theta-\mu)] dF(\theta)=\cos(p\mu)+\int_{0}^{2\pi} p \{[\sin[p(\theta-\mu)]\} F(\theta) d\theta
\end{equation*}
and
\[
\mathbb{E}\{[\sin[p(\Theta-\mu)]\}=\int_{0}^{2\pi} \sin[p(\theta-\mu)] dF(\theta)=-\sin(p\mu)-\int_{0}^{2\pi} p \{[\cos[p(\theta-\mu)]\} F(\theta) d\theta
\]

When $p=1$, applying the Theorem \ref{theorem1}, the EC first moment is given by 
\[
\mu_{1}=\mathbb{E}\{[\cos[(\Theta-\mu)]\}+i\mathbb{E}\{[\sin[(\Theta-\mu)]\}
=\alpha_1+i\beta_1,
\]
where $\alpha_1$ and  $\beta_1$ are determined at the following corollary.

\begin{cor}
	Let $\Theta\sim EC(\beta,\rho,\mu)$. The components of the first central trigonometric moment are given by
	
	\begin{equation*}
	\alpha_1=\cos(\mu)+  \sum_{k=0}^{+\infty}\sum_{s=0}^{k}T_{k,s}
	\left\{A(\beta-k,0,k-s+1) M_0+[\sin(\mu)]^{k-2s}A(\beta-k,0,s+1)M_1\right\}
	\end{equation*}
	and
	\begin{equation*}
	\beta_1=-\sin(\mu)-  \sum_{k=0}^{+\infty}\sum_{s=0}^{k}T_{k,s}\left\{A(\beta-k,1,k-s)M_0+[\sin(\mu)]^{k-2s}A(\beta-k,1,s)M_1\right\},
	\end{equation*}
	where 
	\[
	A(a,b,c)=\int_{0}^{2\pi}\theta^{a}[\cos(\theta-\mu)]^{b}[\sin(\theta-\mu)]^{c} d\theta.
	\]
	\label{cor1}
\end{cor}

Detailed calculations for derivation of the above expressions are presented in Appendix B. To illustrate the use of the Corollary 3.2, for the C distribution ($\beta=1$, $A(0, 0, 1)=A(0,1,1)=A(0,1,0)=0$, $M_0=0$ and $M_1=1$ or $M_0=0$ and $M_1=1$), it
follows that
\begin{align*}
\alpha_1=&\cos(\mu)+T_{0,0}\{A(1,0,1)M_0+A(1,0,1)M_1\}+T_{1,0}\{A(0,0,2)M_0\\
&+\sin(\mu)^{-1}A(0,0,1)M_1\}+T_{1,1}\{A(0,0,1)M_0 +\sin(\mu)^{-1}A(0,0,2)M_1\} \\
=&\cos(\mu)+(2\pi)^{-1}\{-2\pi\cos(\mu)(M_0+M_1)\}+\frac{\rho}{\pi}\pi M_0+\frac{\rho\sin(\mu)}{\pi}[\pi\sin(\mu)^{-1}]M_1\\
=&\rho
\end{align*}
and
\begin{align*}
\beta_1=&-\sin(\mu)-T_{0,0}\{A(1,1,0)M_0+A(1,1,0)M_1\}-T_{1,0}\{A(0,1,1)M_0\\
&+\sin(\mu)A(0,1,0)M_1\}-T_{1,1}\{A(0,1,0)M_0+\sin(\mu)^{-1}A(0,1,1)M_1\}\\
=&-\sin(\mu)+(2\pi)^{-1}\{2\pi\sin(\mu)(M_0+M_1)\}\\
=&0,
\end{align*}
which corresponds to the components of the first central trigonometric moment of the C distribution [\cite{Fisher}]. In a similar manner, the second moment is
\[
\mu_{2}=\mathbb{E}\{[\cos[2(\Theta-\mu)]\}+i\mathbb{E}\{[\sin[2(\Theta-\mu)]\}=\alpha_2+i\beta_2,
\]
where $\alpha_2$ and $\beta_2$ are determined as follows.

\begin{cor} Let $\Theta\sim EC(\beta,\rho,\mu)$. The components of the second central trigonometric moment are given by
	\begin{align*}
	\alpha_2=&\cos(2\mu)+ 4\sum_{k=0}^{\infty}\sum_{s=0}^{k}T_{k,s}\{A(\beta-k,1,k-s+1)M_0\\
	&+[\sin(\mu)]^{k-2s}A(\beta-k,1,s+1)M_1\}
	\end{align*}
	and
	\begin{align*}
	\beta_2=&-\sin(2\mu)-  2\sum_{k=0}^{\infty}\sum_{s=0}^{k}T_{k,s}\{[2A(\beta-k,2,k-s)-A(\beta-k,0,k-s)] M_0\\
	&+
	[\sin(\mu)]^{k-2s}[2A(\beta-k,2,s)-A(\beta-k,0,s)]M_1\}.
	\end{align*}
	\label{cor2}
\end{cor}

\begin{table}[h]
	\centering
	\caption{Ranges of some standard circular measures of the EC distribution.}
	{\begin{tabular}{ccc} 
			\midrule 
			Measure & Expression & Range \\  \midrule
			Mean Resultant Length & $\rho_1$  & $[0,1]$ \\
			\midrule
			Circular Variance & $1-\rho_1$  & $[0,1]$ \\     
			\midrule
			Circular Standard Deviation & $\sqrt {-2\log\rho_1}$ & $[0,\infty)$ \\ 
			\midrule 
			Circular Dispersion & $(1-\alpha_2)/(2{\rho_1}^2)$  & $[0,\infty)$  \\     
			\midrule
			Circular Skewness & $\beta_2/(1-\rho_1)^\frac{3}{2}$ &  $(-\infty,\infty)$ \\     
			\midrule
			Circular Kurtosis & $(\alpha_2-{\rho_1}^4)/(1-\rho_1)^2$ & $(-\infty,\infty)$  \\     
			\midrule
		\end{tabular}}
		\label{range}
	\end{table}

Some standard circular measures are functions of the first and second trigonometric moments. 
The second column of Table \ref{range} presents expressions (in terms of $\rho_1$, $\alpha_2$ and $\beta_2$) for mean resultant length, circular variance, standard deviation, dispersion, skewness and kurtosis of any circular model. Using results of Corollaries \ref{cor1} and \ref{cor2}, these quantities can be obtained to the EC model. The ranges to each resulting EC quantities are given in the third column of Table \ref{range}. The used notation is the same as in \cite{Fisher}, where $\rho_i=\sqrt{\alpha_i^2+\beta_i^2}$ for $i=1,2$. 

\begin{figure}[h]
\centering
{
\resizebox*{12cm}{!}{\includegraphics{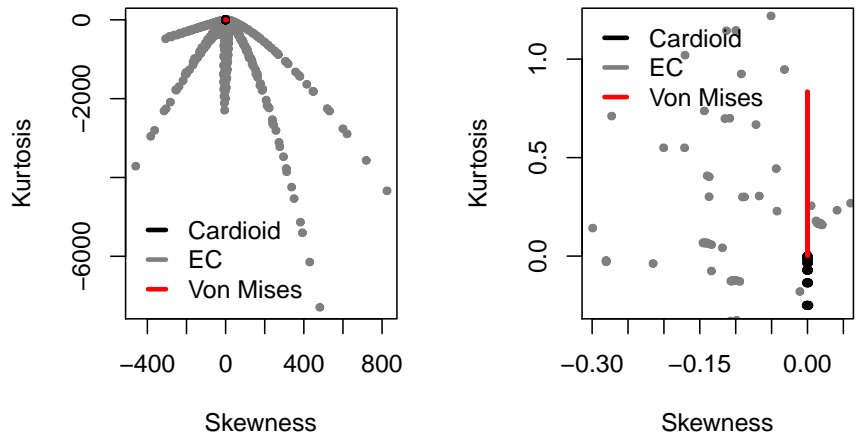}}}
\caption{Skewness and kurtosis maps for EC, C and von Mises distributions.
} \label{skew}
\end{figure}

In order to illustrate the presented measures, Figure \ref{skew} displays the plots of skewness and kurtosis of C, EC and von Mises distributions, which characterize the distribution shape. It may be observed the C (skewness, kurtosis) pair is overlapped to that of the von Mises law and the EC pair region covers the two first. From perspective of this diagram, our model seems to extend the former models. Additionally, C and von Mises skewnesses assume null values, as expected.

\section{Estimation Procedures}
	
\subsection{Maximum Likelihood Estimation}
	
Let $\theta_1, \theta_2, \dots, \theta_n$ be a $n$-points observed sample from $\Theta\sim\text{EC}(\beta,\rho,\mu)$. Then the log-likelihood function at {\boldmath $\delta$}$=(\beta,\rho,\mu)^\top$ is given by
\begin{align*}
l(\mbox{\boldmath$\delta$})=&n\log\beta+(\beta-1)\sum_{i=1}^{n}\log\left\{\frac{\theta_i}{2\pi}+ \frac{\rho}{\pi} [\sin(\theta_i-\mu)+\sin\mu]\right\}\\
&-n\log(2\pi)+\sum_{i=1}^{n}\log\{1+2 \rho \cos(\theta_i-\mu)\}.
	\end{align*}
Therefore, the ML estimates for $\beta$, $\rho$ and $\mu$, say $\hat{\beta}$, $\hat{\rho}$ and $\hat{\mu}$, can be defined as solutions of the following non-linear system:
	\begin{equation}
	\frac{\partial l(\mbox{\boldmath$\delta$})}{\partial \beta}\Bigg|_{\mbox{\boldmath$\delta$}=\mbox{\boldmath$\hat{\delta}$}}=\frac{n}{\beta}+\sum_{i=1}^{n}\log\left\{\frac{\theta_i}{2\pi}+\frac{\rho}{\pi}[\sin(\theta_i-\mu)+\sin\mu]\right\}=0,
	\label{mv1}
	\end{equation}
	\begin{equation}
	\frac{\partial l(\mbox{\boldmath$\delta$})}{\partial \rho}\Bigg|_{\mbox{\boldmath$\delta$}=\mbox{\boldmath$\hat{\delta}$}}=\sum_{i=1}^{n}
	\left\{(\beta-1)\frac{\sin(\theta_i-\mu)+\sin\mu}{\theta_i+2\rho[\sin(\theta_i-\mu)+\sin\mu]}+\frac{\cos(\theta_i-\mu)}{1+2\rho\cos(\theta_i-\mu)}\right\}=0
	\label{mv2}
	\end{equation}
	and
	\begin{equation}
	\frac{\partial l(\mbox{\boldmath$\delta$})}{\partial \mu}\Bigg|_{\mbox{\boldmath$\delta$}=\mbox{\boldmath$\hat{\delta}$}}=\sum_{i=1}^{n}
	\left\{(\beta-1)\frac{-\cos(\theta_i-\mu)+\cos\mu}{\theta_i+2\rho[\sin(\theta_i-\mu)+\sin\mu]}+\frac{\sin(\theta_i-\mu)}{1+2\rho\cos(\theta_i-\mu)}\right\}=0.
	\label{mv3}
	\end{equation}
	This system can be reduced to others under equations (\ref{mv2}) and (\ref{mv3}), replacing the ML estimate $\hat{\beta}$ by
	\begin{equation}
	\hat{\beta}(\rho,\mu)=-\frac{n}{\sum_{i=1}^{n}\log\left\{\frac{\theta_i}{2\pi}+\frac{\rho}{\pi}[\sin(\theta_i-\mu)+\sin\mu]\right\}},
	\label{betahat}
	\end{equation}
	which is obtained from (\ref{mv1}). Thus, the ML estimates for ${\rho}$ and ${\mu}$ are obtained numerically from $\frac{\partial l(\mbox{\boldmath$\delta$})}{\partial \rho}\Bigg|_{\mbox{\boldmath$\delta$}=\mbox{\boldmath$\hat{\delta}$}}=0$ and $\frac{\partial l(\mbox{\boldmath$\delta$})}{\partial \mu}\Bigg|_{\mbox{\boldmath$\delta$}=\mbox{\boldmath$\hat{\delta}$}}=0$.

	\subsection{Quantile Least Squares Method}
	
	The QLSE for {\boldmath$\delta$} can be defined as solutions from minimization of the sum of squares of the differences between theoretical and empirical quantiles. Consider $\theta_{1:n},\cdots,\theta_{n:n}$ as observed order statistics drawn from $n-$points random sample of $\Theta\sim EC(\beta,\rho,\mu)$, where $\theta_{k:n}$ is the $k$th order statistics. Thus, the QLS estimates for EC parameters consist in argument that minimizes the following goal function:
	
	\begin{equation}
	q(\mbox{\boldmath$\delta$})=\sum_{i=1}^{n}\left[\frac{i}{n}-\left\{\frac{\theta_{i:n}}{2\pi}+\frac{\rho}{\pi}[\sin(\theta_{i:n}-\mu)+\sin(\mu)]\right\}^{\beta}\right]^{2}.
	\end{equation}
	Equivalently to discussed in previous section, the QLS estimates for $\beta$, $\rho$ and $\mu$ can be defined as solutions of the following non-linear equations system:
	\begin{equation}
	\frac{\partial q(\mbox{\boldmath$\delta$})}{\partial \beta}=\sum_{i=1}^{n}\left[\frac{i}{n}-F(\theta_{i:n})\right]F_C(\theta_{i:n})^{2\beta-1}\log[F_C(\theta_{i:n})]=0,
	\end{equation}
	\begin{equation}
	\frac{\partial q(\mbox{\boldmath$\delta$})}{\partial \rho}=\sum_{i=1}^{n}\left[\frac{i}{n}-F(\theta_{i:n})\right]F_C(\theta_{i:n})^{\beta-1}\frac{1}{\pi}[\sin(\theta_{i:n}-\mu)+\sin\mu]=0
	\end{equation}
	and
	\begin{equation}
	\frac{\partial q(\mbox{\boldmath$\delta$})}{\partial \mu}=\sum_{i=1}^{n}\left[\frac{i}{n}-F(\theta_{i:n})\right]F_C(\theta_{i:n})^{\beta-1}\frac{\rho}{\pi}[\cos(\theta_{i:n}-\mu)-\cos\mu]=0,
	\end{equation}
	where $F_C$ represents the C cdf and $F$ is given in (\ref{acum}).  
	
\section{Numerical Results}
	
\subsection{Simulation Study}
	
First, a Monte Carlo simulation study was performed to assess and compare the two proposed estimators. To that end, five thousand replications were considered and, on each one of them, bias and mean square error for both procedures were quantified, like comparison criteria.
	
\begin{table}[htp]
\centering
\caption{MSEs for the ML estimates using as estimation system (\ref{mv1})-(\ref{mv3}) labelled as ``MSEa" and (\ref{betahat}) in (\ref{mv2})-(\ref{mv3}) denoted as ``MSEb" are considered. The size of the sample and Monte Carlo replications were set to $100$ and $5.000$, respectively.}
{\begin{tabular}{ccc} 
\hline 
\multirow{1}{3.2cm}{\centering{$(\beta,\rho,\mu)$}} & \multirow{1}{4.2cm}{\centering{MSEa}} & \multirow{1}{4.2cm}{\centering{MSEb}} \\
  \hline 
\multirow{1}{3.2cm}{\centering{$\left(1,\frac{1}{2},2\pi\right)$}}
& 
\multirow{1}{4.2cm}{\centering{$\left(0.0071,0.2991,0.0000\right)$}} & \multirow{1}{4.2cm}{\centering{$\left(0.0000,0.0000,0.0000\right)$}}\\
\hline 
\multirow{1}{3.2cm}{\centering{$\left(4,\frac{3}{10},\frac{2\pi}{3}\right)$}} & 
\multirow{1}{4.2cm}{\centering{$\left(0.6449,13.7865,0.0764\right)$}} & \multirow{1}{4.2cm}{\centering{$\left(0.5101,0.0030,0.0516\right)$}} \\
\hline  
\multirow{1}{3.2cm}{\centering{$\left(1,\frac{3}{10},\frac{4\pi}{3}\right)$}} & 
\multirow{1}{4.2cm}{\centering{$\left(0.0658,0.5818,0.1883\right)$}} & \multirow{1}{4.2cm}{\centering{$\left(0.0293,0.0073,0.0936\right)$}}\\
\hline 
\multirow{1}{3.2cm}{\centering{$\left(4,\frac{1}{2},\frac{\pi}{3}\right)$}} & 
\multirow{1}{4.2cm}{\centering{$\left(0.2384,11.4078,0.0119\right)$}} & \multirow{1}{4.2cm}{\centering{$\left(0.1761,0.0001,0.0056\right)$}}\\
\hline 
\end{tabular}}
\label{23parameters}
\end{table}
		
Initially, a discussion about the effect of using (6) is presented in estimation process by maximum likelihood. Here, a sample size $n=100$ and four parametric points are considered. Results are displayed in Table \ref{23parameters} and indicate that the use of (6) may imply in more accurate estimates. The most pronounced improvement can be observed to estimate $\rho$. Moreover, the estimation considering (6) takes a mean execution of $83$ seconds, while the other is slower, taking $116$ seconds. From now on, the best ML estimates for $\beta$, $\rho$ and $\mu$ will be used.

\newpage
		
\begin{table}[htp]
\footnotesize
\centering
\caption{Average bias e average MSE for the ML and QLS estimates for different values of $\beta$, $\rho$ and $\mu$, by the Monte Carlo method, over $5.000$ replications. The size of the sample was set to $30$, $50$ and $100$. The parametric vectors of the first column are sorted by mean direction.}
{\begin{tabular}{ccccc} 
\hline
$(\beta,\rho,\mu)$ & Method &  $n$ & Bias & MSE \\  \hline
\multirow{6}{2cm}{\centering{$\left(1,\frac{1}{2},2\pi\right)$}} & \multirow{3}{1.4cm}{\centering{MLE}} & $30$ &  $\left(0.0315,0.0000,0.0000\right)$ & $\left(0.0272,0.3683,0.0000\right)$
\\ 
&& $50$ & 
$\left(0.0170,0.0000,0.0000\right)$ & $\left(0.0000,0.0000,0.0000\right)$
\\ 
&& $100$ & 
$\left(0.0094,0.0000,0.0000\right)$ & $\left(0.0000,0.0000,0.0000\right)$ \\
\cline{2-5}    
& \multirow{3}{1.4cm}{\centering{QLSE}} & $30$ & 
$\left(0.0685,0.0000,0.0000\right)$ & $\left(0.0213,0.3398,0.0000\right)$\\ 
&& $50$ & 
$\left(0.0546,0.0000,0.0000\right)$ & $\left(0.0129,0.3176,0.0000\right)$\\ 
&& $100$ & 
$\left(0.0421,0.0000,0.0000\right)$ & $\left(0.0071,0.2991,0.0000\right)$\\ 
\hline

\multirow{6}{2cm}{\centering{$\left(1,\frac{3}{10},\frac{4\pi}{3}\right)$}} & \multirow{3}{1.4cm}{\centering{MLE}} & $30$ & 
$\left(0.1343,0.0128,-0.1473\right)$ & $\left(0.2093,0.0173,0.5766\right)$\\ 
&& $50$ & 
$\left(0.0703,0.0011,-0.0873\right)$ & $\left(0.0842,0.0126,0.2838\right)$   \\ 
&& $100$ & 
$\left(0.0330,-0.0030,-0.0361\right)$ & $\left(0.0293,0.0073,0.0936\right)$\\
\cline{2-5}    
& \multirow{3}{1.4cm}{\centering{QLSE}} & $30$ & 
$\left(0.0823,0.0419,-0.1351\right)$ & $\left(0.3230,0.9282,0.5750\right)$\\
&& $50$ & 
$\left(0.0337,0.0254,-0.1035\right)$ & $\left(0.1420,0.6792,0.3726\right)$\\
&& $100$ & 
$\left(0.0185,0.0104,-0.0679\right)$ & $\left(0.0658,0.5818,0.1883\right)$\\
\hline
\multirow{6}{2cm}{\centering{$\left(4,\frac{1}{2},\frac{\pi}{3}\right)$}} & \multirow{3}{1.4cm}{\centering{MLE}} & $30$ & 
$\left(0.0683,-0.0089,-0.0067\right)$ & $\left(0.6570,0.0006,0.0215\right)$\\ 
&& $50$ & 
$\left(0.0249,-0.0068,-0.0045\right)$ & $\left(0.3651,0.0003,0.0123\right)$ \\ 
&& $100$ & 
$\left(0.0112,-0.0041,-0.0025\right)$ & $\left(0.1761,0.0001,0.0056\right)$\\
\cline{2-5}    
& \multirow{3}{1.4cm}{\centering{QLSE}} & $30$ & 
$\left(-0.3654,-0.0311,-0.0570\right)$ & $\left(1.0033,10.6952,0.0581\right)$\\
&& $50$ & 
$\left(-0.2616,-0.0236,-0.0362\right)$ & $\left(0.4815,10.9000,0.0293\right)$\\
&& $100$ & 
$\left(-0.1544,-0.0165,-0.0191\right)$ & $\left(0.2384,11.4078,0.0119\right)$\\
\hline
\end{tabular}}
\label{simulation1}
\end{table}

\begin{table}[htp]
\centering
\footnotesize
\caption{Average bias e MSE for the ML and QLS estimates for different values of $\beta$, $\rho$ and $\mu$, by the Monte Carlo method, over $5.000$ replications. The size of the sample was set to $30$, $50$ and $100$. The parametric vectors of the first column are sorted by mean resultant length.}
{\begin{tabular}{ccccc} 
\hline
$(\beta,\rho,\mu)$ & Method &  $n$ & Bias & MSE \\  \hline
\multirow{6}{2cm}{\centering{$\left(0.3,\frac{1}{2},\frac{4\pi}{3}\right)$}} & \multirow{3}{1.4cm}{\centering{MLE}} & $30$ &  $\left(0.0131,-0.0076,0.0135\right)$ & $\left(0.0041,0.0007,0.0242\right)$ 
\\ 
&& $50$ & 
$\left(0.0069,-0.0049,0.0017\right)$ & $\left(0.0021,0.0002,0.0113\right)$  
\\ 
&& $100$ & 
$\left(0.0036,-0.0029,-0.0035\right)$ & $\left(0.0010,0.0001,0.0048\right)$\\ 
\cline{2-5}  
& \multirow{3}{1.4cm}{\centering{QLSE}} & $30$ & 
$\left(0.0064,-0.0219,-0.0045\right)$ & $\left(0.0098,0.0185,0.0739\right)$\\
						&& $50$ & 
						$\left(0.0038,-0.0156,0.0017\right)$ & $\left(0.0035,0.0128,0.0378\right)$\\ 
						&& $100$ & 
						$\left(0.0036,-0.0111,0.0043\right)$ & $\left(0.0012,0.0105,0.0191\right)$\\ 
						\hline
						
						\multirow{6}{2cm}{\centering{$\left(4,\frac{3}{10},\frac{\pi}{3}\right)$}} & \multirow{3}{1.4cm}{\centering{MLE}} & $30$ &   $\left(0.3223,0.0178,0.0031\right)$ & $\left(1.5698,0.0096,0.1673\right)$   
						\\ 
						&& $50$ & 
						$\left(0.1696,0.0106,-0.0062\right)$ & $\left(0.7629,0.0060,0.0969\right)$    
						\\ 
						&& $100$ & 
						$\left(0.0852,0.0050,-0.0052\right)$ & $\left(0.3411,0.0032,0.0440\right)$\\ 
						\cline{2-5}  
						& \multirow{3}{1.4cm}{\centering{QLSE}} & $30$ & 
						$\left(-0.0977,0.0034,-0.1103\right)$ & $\left(3.2670,15.5235,0.3313\right)$\\
						&& $50$ & 
						$\left(-0.1147,0.0027,-0.1092\right)$ & $\left(1.1496,13.9908,0.1954\right)$\\
						&& $100$ & 
						$\left(-0.0603,0.0006,-0.0639\right)$ & $\left(0.5405,13.7842,0.0832\right)$\\ 
						\hline
						
						\multirow{6}{2cm}{\centering{$\left(10,\frac{1}{2},2\pi\right)$}} & \multirow{3}{1.4cm}{\centering{MLE}} & $30$ &     $\left(0.3154,0.0000,0.0000\right)$ & $\left(3.9682,0.0000,0.0000\right)$ 
						\\ 
						&& $50$ & 
						$\left(0.1698,0.0000,0.0000\right)$ & $\left(2.1771,0.0000,0.0000\right)$\\    
						&& $100$ & 
						$\left(0.0935,0.0000,0.0000\right)$ & $\left(1.0645,0.0000,0.0000\right)$\\  
						\cline{2-5}  
						& \multirow{3}{1.4cm}{\centering{QLSE}} & $30$ & 
						$\left(0.3048,0.0000,0.0000\right)$ & $\left(0.2787,96.3191,0.0000\right)$\\
						&& $50$ & 
						$\left(0.3012,0.0000,0.0000\right)$ & $\left(0.2685,96.2410,0.0000\right)$\\ 
						&& $100$ & 
						$\left(0.2861,0.0000,0.0000\right)$ & $\left(0.2405,95.9267,0.0000\right)$\\ 
						\hline
					\end{tabular}}
					\label{simulation2}
				\end{table}

			Now, methods discussed in previous sections are compared. Results are presented in Tables \ref{simulation1} and \ref{simulation2}. They are indexed in terms of mean direction and resultant lenght under ascending order for $n=30$, $50$ and $100$. In regard to choose parametric scenarios, vectors $(\beta,\rho,\mu)$ are selected such that they have different mean directions and resultant lengths. 
			
\begin{figure}[h]
\centering
{
\resizebox*{7cm}{!}{\includegraphics{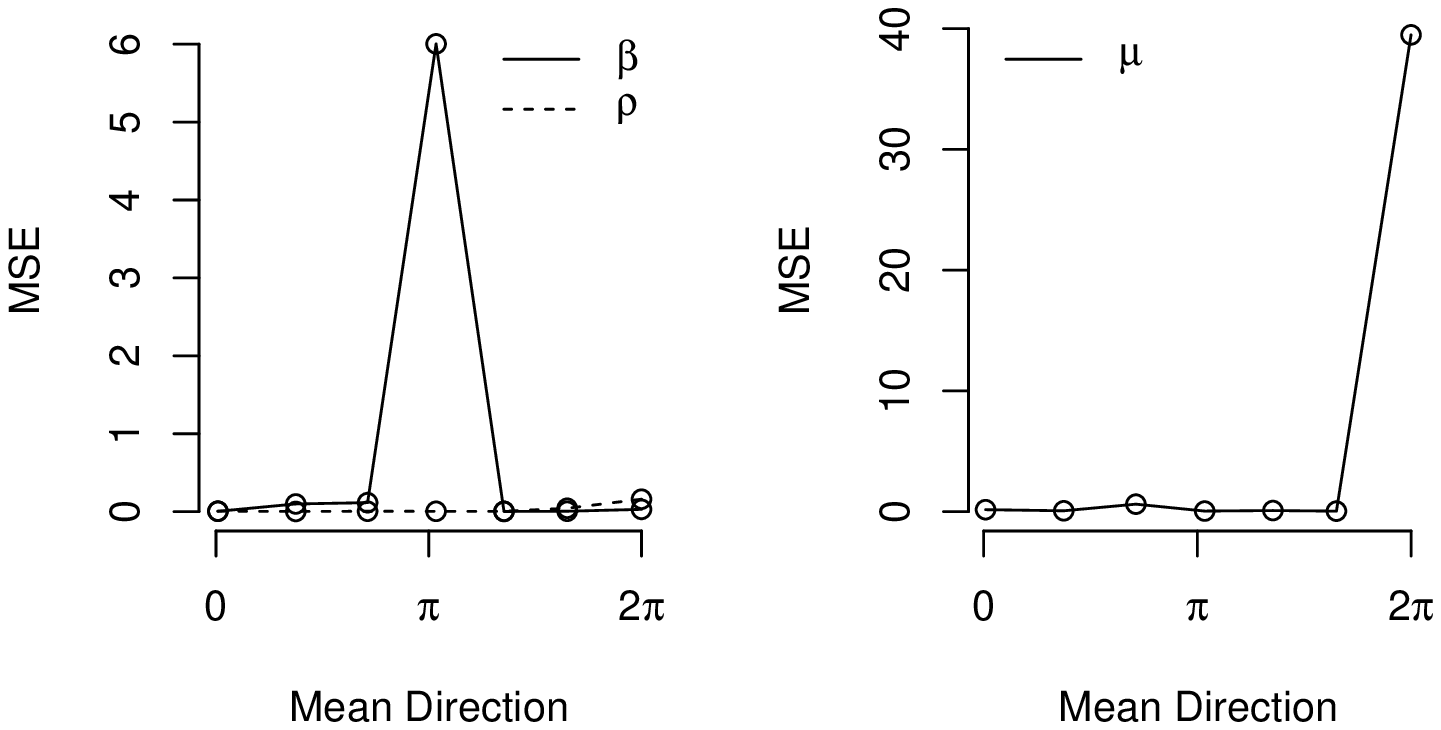}}}
\caption{
MSE for the ML estimates of the seven parametric vectors chosen according to the mean direction.} \label{mse1}
\end{figure}

\begin{figure}[h]
\centering
{
\resizebox*{7cm}{!}{\includegraphics{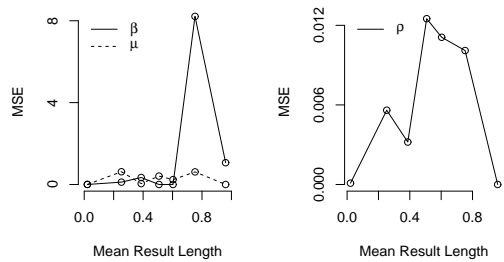}}}
\caption{MSE for the ML estimates of the seven parametric vectors chosen according to the mean resultant length.} \label{mse2}
\end{figure}

\begin{figure}[h]
\centering
{
\resizebox*{7cm}{!}{\includegraphics{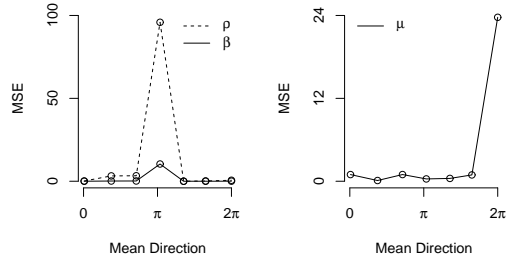}}}
\caption{MSE for the QLS estimates of the seven parametric vectors chosen according to the mean direction.
} \label{mse11}
\end{figure}

\begin{figure}[h]
\centering
{
\resizebox*{7cm}{!}{\includegraphics{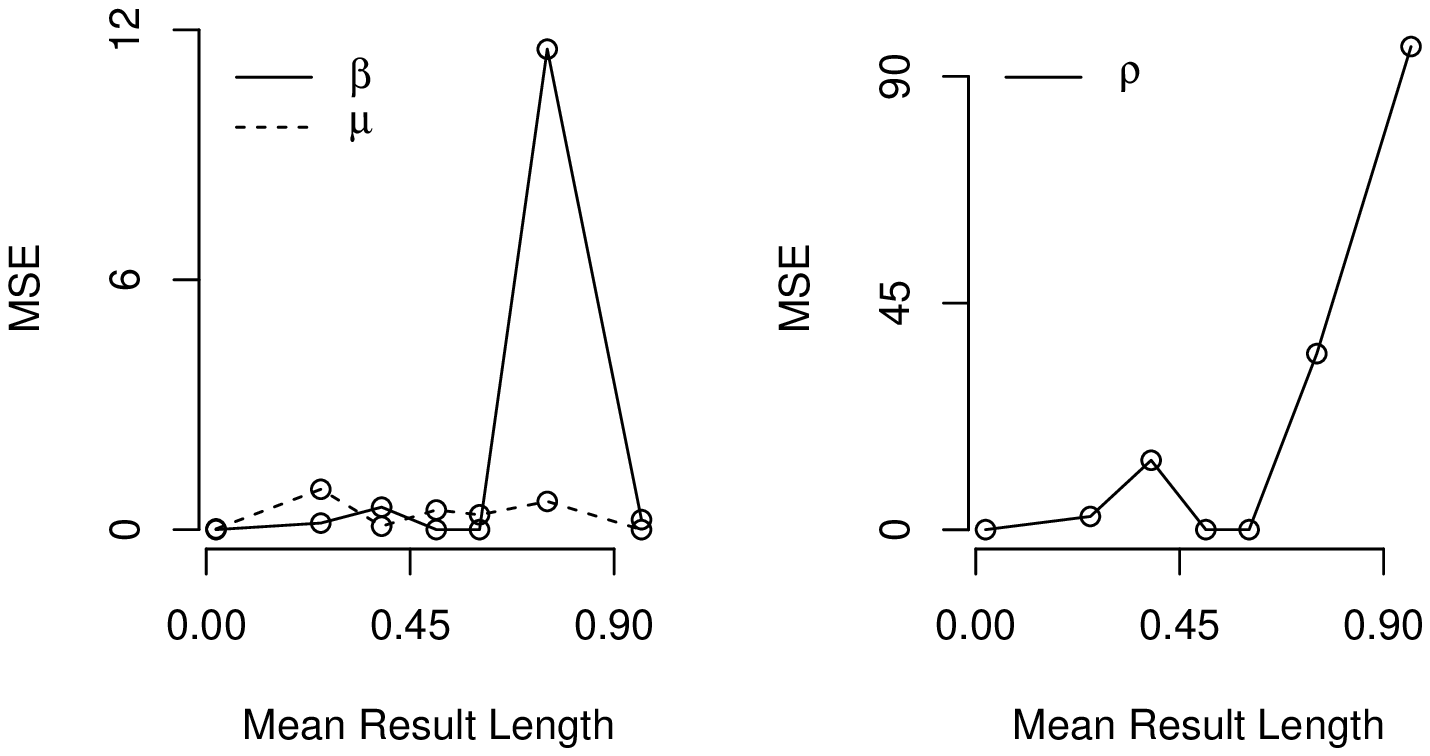}}}
\caption{MSE for the QLS estimates of the seven parametric vectors chosen according to the mean resultant length.} 
\label{mse22}
\end{figure}

			Figures \ref{mse1} and \ref{mse11} show the MSEs for the ML and QLS estimates at parameter vectors of Table \ref{simulation1} added to other four, respectively. Figures \ref{mse2} and \ref{mse22} also address MSEs for points in Table \ref{simulation2} along with ones. To illustrate the abscissas of graphs, Figure \ref{mse1} begins with $\left(0.6,0.2,\frac{\pi}{6}\right)$ with mean direction $0.0279$ and, after, by $(2,0.3,\frac{2\pi}{3})$ with mean direction $1.1777$.
			
\newpage
			
In regard to variation of mean direction, fourth and seventh points impose difficulties to both methods for estimating the parameters. However, the impact over ML estimates are smaller than that on the QLS estimates. For the variation of mean resultant length, the hardest scenario is the sixth point and the same conclusion is obtained. In particular, poor QLS estimates for $\rho$ at third, sixth and seventh points are found, according to Figure \ref{mse22}, in contrast with respective ML estimates. Additionally, such points refer to $\beta=4$, $\beta=4$ and $\beta=10$, respectively, which indicates that high values to $\beta$ difficult the estimation of $\rho$. 
			
Interesting evidence can be found in Figure \ref{mse22}, where the behavior of MSEs for estimates of $\rho$ is approximately monotone.
			
In general, the MLE performed better than the QLSE in most considered cases, comparing the MSEs of the estimates. The QLSE seems to be more indicated for large $\beta$ ($\beta = 10$) and small $\rho$ ($\rho=\frac{1}{5}$) (see Table \ref{mleqlse}).

\newpage
								
\begin{table}[h]
	\centering
\caption{Average EQM for the ML and QLS estimates for different values of $\beta$, $\rho$ and $\mu$, by the Monte Carlo method, over $5.000$ replications and sample size $n=100$.}
{\begin{tabular}{ccc} 
\hline 
\multirow{1}{3.2cm} {\centering$(\beta,\rho,\mu)$} & \multirow{1}{4.2cm}{\centering ML} & \multirow{1}{4.2cm}{\centering QLS} \\  \hline \vspace{-0.1cm} 
\multirow{1}{3.2cm}{\centering{$\left(0.3,\frac{1}{5},\frac{2\pi}{3}\right)$}} 
& 
\multirow{1}{4.2cm} {\centering $\left(0.0014,0.0125,0.4106\right)$} &   \multirow{1}{4.2cm}{\centering $\left(0.0019,0.0104,0.4729\right)$}\\     
  \hline
\multirow{1}{3.2cm}{\centering{$\left(0.3,\frac{1}{5},\frac{\pi}{6}\right)$}}    
& 
\multirow{1}{4.2cm}{\centering$\left(0.0011,0.0111,0.2384\right)$} &   \multirow{1}{4.2cm}{\centering$\left(0.0013,0.0106,0.3574\right)$}\\  \hline  
\multirow{1}{3.2cm}{\centering{$\left(10,\frac{1}{2},2\pi\right)$}}  
& 
\multirow{1}{4.2cm}{\centering$\left(1.0645,0.0000,0.0000\right)$} &   \multirow{1}{4.2cm}{\centering$\left(0.2405,95.9267,0.0000\right)$}\\     \hline 
\end{tabular}}
\label{mleqlse}
\end{table}

\subsection{Application}
			
In order to illustrate the potentiality of the EC distribution, an application to real data was made. Further, its performance was compared with other due to the Cardioid and von Mises models. ML estimates were used to fit considered models to data. All the computations were done using function \texttt{maxLik} at the R statistical software [\cite{R}].
					
\begin{figure}[h]
\centering
{
\resizebox*{4.5cm}{3.8cm}{\includegraphics{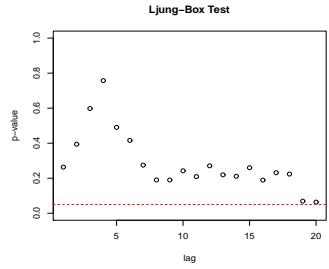}}} 
\caption{Ljung-Box test statistic.} \label{ind}
\end{figure}

The dataset consists of $21$ wind directions obtained by a Milwaukee weather station, at 6.00 am on consecutive days [\cite{Data}] (see Appendix C). The independence of the data can be verified by the Box-Pierce (Ljung-Box) test [\cite{Independence}] in Figure \ref{ind}.
					
\begin{figure}[h]
\centering
{
\resizebox*{5.5cm}{4.2cm}{\includegraphics{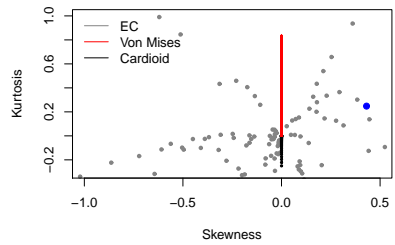}}} 
\caption{Data skewness and kurtosis.} \label{skewdados}
\end{figure}

The Figure \ref{skewdados} shows the sample skewness and kurtosis of data, by the blue point given by $0.4313$ and $0.2480$, respectively. It noticeable that the EC model may provide better fit than those due to C and von Mises distributions. The likelihood ratio test was also applied to compare the C ($H_0: \beta = 1$) and EC ($H_0: \beta \neq 1$) distributions. The p-value is $ 0.0027 $, indicating the EC model as the best descriptor for these wind directions. In what follows, other quantitative discussions are done.
					
First, the ML estimates and their SEs (given in parentheses) are evaluated and, subsequently, the values of the Kuiper (K) and Watson (W) statistics are obtained. These aderence measures may be found in \cite{Jama} and they are given by
\[
K=\sqrt n\left\{\max_{\substack{1\leq i \leq n}} \left(U_{(i)}-\frac{i-1}{n}\right)+\max_{\substack{1\leq i \leq n}}\left(\frac{i}{n}-U_{(i)}\right)\right\}
\]
and
\[
W=\sum_{i=1}^{n}\left[ \left(U_{(i)}-\frac{i-0.5}{n}\right)-\left(\overline{U}-0.5\right)\right]^2+\frac{1}{12n},
\]
where $U_{(i)}=F(\alpha_{(i)})$ in terms of the ordered observations $\alpha_{(1)} \leq \alpha_{(2)} \leq \cdots \leq \alpha_{(n)}$. In general, smaller values of them are associated to better fits. Table \ref{aju} displays results from where we concluded that the EC distribution could be chosen as the best model for the data set. 
					
\begin{figure}[h]
\centering
\subfigure{
\resizebox*{6.1cm}{6.1cm}{\includegraphics{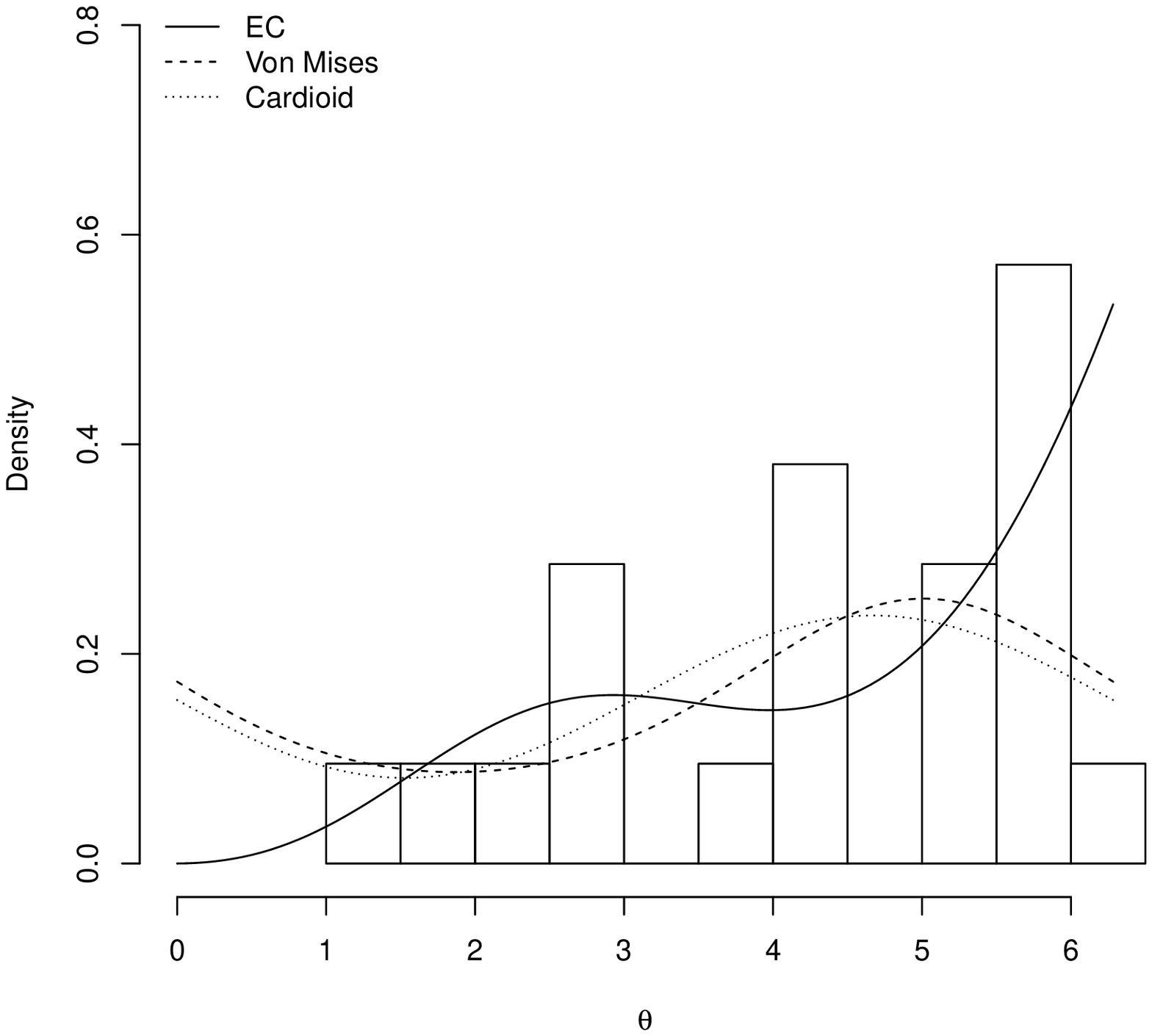}}}
\centering
\subfigure{
\resizebox*{6.1cm}{6.1cm}{\includegraphics{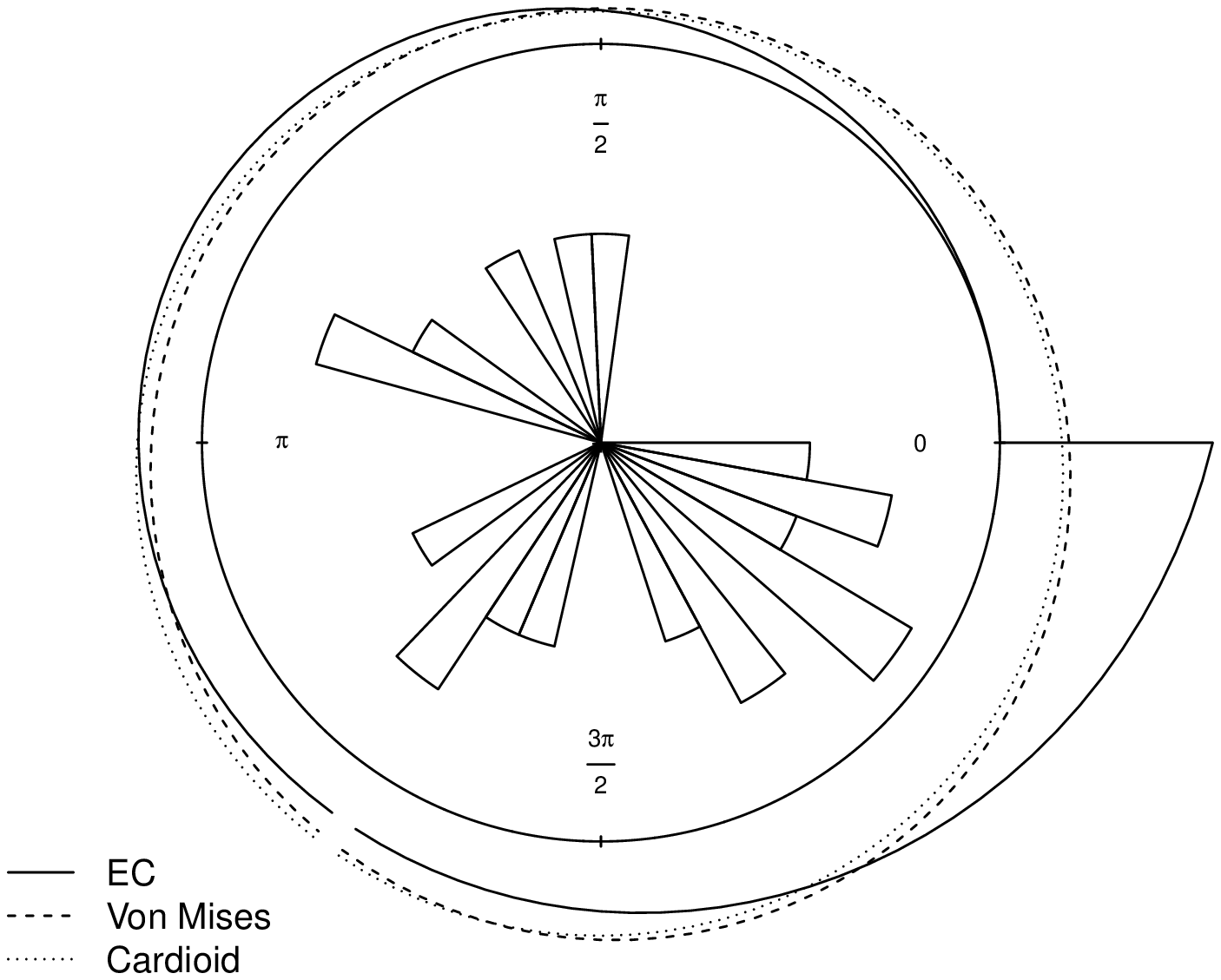}}}
\caption{Fitted densities of the EC, C and von Mises models for the real data.} \label{ajuste}
\end{figure}

In order to do a qualitative comparison, Figure \ref{ajuste} presents empirical and fitted densities. Results confirm what is concluded from Table \ref{aju}. 
					
\begin{table}[h]
\centering
\caption{ML estimates of the model parameters for the data, the corresponding standard errors (given in parentheses) and the Kuiper and Watson statistics.}
{\begin{tabular}{ccccccc} 
\midrule 
Model &  $\beta$ & $\rho$ & $\mu$  & Kuiper &  Watson \\  \midrule
\multirow{2}{5cm}{\centering{Cardioid}} & $-$ & $0.2436$  & $4.6708$  & \multirow{2}{1cm}{\centering{$1.0388$}} & \multirow{2}{1cm}{\centering{$0.0592$}}\\ 
& $-$ & ($0.1463$) & ($0.6835$) & & \\     
\midrule
\multirow{2}{5cm}{\centering{Exponentiated Cardioid}} & $2.8757$ & $0.2164$ & $1.1782$  & \multirow{2}{1cm}{\centering{$0.7369$}} & \multirow{2}{1cm}{\centering{$0.0257$}}\\ 
& $(0.8929)$ & $(0.1465)$ & $(0.6168)$  &  & \\     
\midrule
\multirow{2}{5cm}{\centering{Von Mises}} & $-$ & $0.5322$ & $5.0092$ & \multirow{2}{1cm}{\centering{$1.1590$}} & \multirow{2}{1cm}{\centering{$0.0711$}}\\ 
& $-$ & ($0.3250$) & ($0.5899$) & & \\   
\hline
\end{tabular}}
\label{aju}
\end{table}

\section{Concluding remarks}
An extended cardioid model has been proposed, called the \emph{Exponentiated Cardioid} distribution. Our proposal has shown to be able to describe circular asymmetric data, as well as amodality, unimodality and bimodality scenarios. Expressions for the EC trigonometric moments by means of expansions were derived and a discussion about its quantile function and mode was presented. Two estimation procedures for the EC parameters were proposed in regard to maximum likelihood and quantile least square frameworks. The performance of these estimates was quantified through a Monte Carlo simulation study. Finally, an application to real wind data was made and results have indicated that the EC model may outperform the classic C and von Mises distributions.
						
\section*{Appendix}

\appendix
						
\section{The proof of Theorem 2.1.}\label{quantilesproof}
						
Let $\Theta\sim EC(\beta,\rho,\mu)$. For $\theta \in [0,0.6]$, $\theta \approx \sin(\theta)$. Thus, $\theta-\mu \approx \sin(\theta-\mu)$, for $\theta-\mu \in [0,0.6]$. In this case, being $\theta=Q_\alpha$, where $Q_\alpha$  is the EC quantil at $\alpha$, it follows from (1): 
						
\begin{align*}
&\alpha=\left\{\frac{Q_\alpha}{2\pi}+\frac{\rho}{\pi}[\sin(Q_\alpha-\mu)+\sin (\mu)]\right\}^{\beta}\Rightarrow \alpha^{\frac{1}{\beta}}=\frac{Q_\alpha}{2\pi}+\frac{\rho}{\pi}[\sin(Q_\alpha-\mu)+\sin (\mu)]\\
& \Rightarrow   \alpha^{\frac{1}{\beta}}\approx \frac{Q_\alpha}{2\pi}+\frac{\rho}{\pi}[Q_\alpha-\mu+\sin (\mu)] \Rightarrow \alpha^{\frac{1}{\beta}}+\frac{\rho}{\pi}[\mu-\sin (\mu)]\approx \frac{Q_\alpha}{2\pi}(1+2\rho)\\
&\Rightarrow 
Q_\alpha\approx \frac{2\pi}{1+2\rho}\left\{\alpha^{\frac{1}{\beta}}+\frac{\rho}{\pi}[\mu-\sin (\mu)]\right\}.
\end{align*}

The items $(3)$ and $(5)$ of this Theorem are demonstrated in a similar way. 
						
Now, let $\Theta\sim EC(\beta,\rho,\mu)$. For $\theta \in [0.6,2.62]$, the quadratic function $-\frac{1}{2}\theta^2+\frac{\pi}{2}\theta+\frac{8-\pi^2}{8} \approx \sin(\theta)$ obtained from the second degree Taylor polynomial for $\sin(\theta)$ around the value $\frac{\pi}{2}$ is used. Thus, $-\frac{1}{2}(\theta-\mu)^2+\frac{\pi}{2}(\theta-\mu)+\frac{8-\pi^2}{8} \approx \sin(\theta-\mu)$, for $\theta-\mu \in [0.6,2.62]$. In this case, 
\begin{align*}
&\alpha=\left\{\frac{Q_\alpha}{2\pi}+\frac{\rho}{\pi}[\sin(Q_\alpha-\mu)+\sin (\mu)]\right\}^{\beta}\Rightarrow \alpha^{\frac{1}{\beta}}=\frac{Q_\alpha}{2\pi}+\frac{\rho}{\pi}[\sin(Q_\alpha-\mu)+\sin (\mu)]\\
& \Rightarrow   \alpha^{\frac{1}{\beta}}\approx \frac{Q_\alpha}{2\pi}+\frac{\rho}{\pi}\left[-\frac{1}{2}(Q_\alpha-\mu)^2+\frac{\pi}{2}(Q_\alpha-\mu)+\frac{8-\pi^2}{8}+\sin (\mu)\right] \\
&\Rightarrow   \alpha^{\frac{1}{\beta}}-\frac{\rho}{\pi}\sin(\mu)-\frac{\pi}{\rho}\left\{\frac{\mu}{2}[-\mu-\pi]+\frac{8-\pi^2}{8}\right\}\approx -\frac{\rho}{2\pi}Q_\alpha^2+\left\{\frac{1}{\pi}\left[\frac{1}{2}+\rho\mu+\frac{\pi\rho}{2}\right]\right\}Q_\alpha.
\end{align*}
						
Let $C=\alpha^{\frac{1}{\beta}}-\frac{\rho}{\pi}\sin(\mu)-\frac{\pi}{\rho}\left\{\frac{\mu}{2}[-\mu-\pi]+\frac{8-\pi^2}{8}\right\}$ and $D=\frac{1}{\pi}\left[\frac{1}{2}+\rho\mu+\frac{\pi\rho}{2}\right]$, then
\[ -\frac{\rho}{2\pi}Q_\alpha^2+DQ_\alpha-C \approx 0
\Rightarrow  
Q_\alpha\approx\frac{\pi}{\rho}[D-\sqrt{E}],
\]
where $E=D^2-\frac{2\rho C}{\pi}$.

The item $4$ of the theorem is demonstrated in a similar manner. In this case, $\frac{1}{2}\theta^2-\frac{3\pi}{2}\theta+\frac{9\pi^2-8}{8}$ is the second degree Taylor polynomial for $\sin(\theta)$ around the value $\frac{3\pi}{2}$.
						
\begin{figure}[h]
\centering
{
\resizebox*{7cm}{!}{\includegraphics{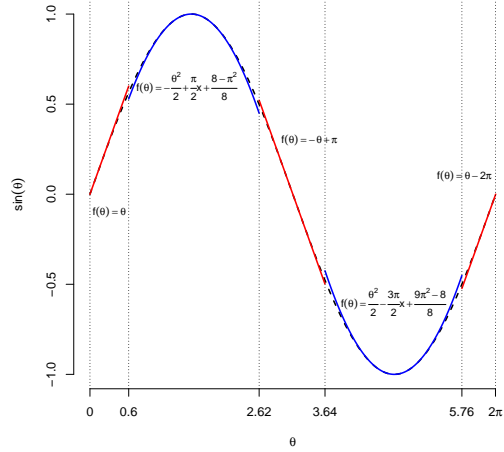}}}\vspace{-0.3cm}
\caption{
Functions that approximate $\sin(\theta)$ and respective intervals of variation.} \label{figquantiles}
\end{figure}

As on illustration, the Figure \ref{figquantiles} shows the plot of $\sin(\theta)$ and the functions used for the approximations.
						
\section{First central trigonometric moment}
In this appendix, the expression for the first central trigonometric moment of the EC distribution is derived in detail.
\begin{thm}[Binomial Theorem] It is known, for $r \in \mathbb{R}$ ou $|\frac{x}{y}|<1$,
\begin{equation*}
(x+y)^r=\sum_{k=0}^{\infty}{r \choose k}x^k y^{r-k}.
\end{equation*}
\end{thm} 
Consider also the following trigonometric inequalities and relation: For $x \in \mathbb{R}$, i) $|\sin(x)|<x$ and $|\cos(x)|<1$ and ii) $\sin(x)+\cos(y)=2\sin\left(\frac{x+y}{2}\right)\cos \left(\frac{x-y}{2}\right)$ hold and, consequently, one has that
\begin{equation*}
\left|\frac{\frac{\rho}{\pi}[\sin(\theta-\mu)+\sin (\mu)]}{\frac{\theta}{2\pi}}\right|=\left|\frac{2\rho[\sin(\theta-\mu)+\sin (\mu)]}{\theta}\right|=\left|\frac{4\rho}{\theta}\sin\left(\frac{\theta}{2}\right)\cos\left(\frac{\theta-2\mu}{2}\right)\right|
\end{equation*}
\begin{equation*}
<\left|\frac{4\rho}{\theta}\frac{\theta}{2}\cos\left(\frac{\theta-2\mu}{2}\right)\right|\leq 2\rho \leq 1.
\end{equation*}
						
Thus, from the binomial theorem, 
the EC cdf can be written as

\begin{equation*}
\left\{\frac{\theta}{2\pi}+\frac{\rho}{\pi}[\sin(\theta-\mu)+\sin (\mu)]\right\}^{\beta}=\sum_{k=0}^{\infty}{\beta \choose k}\left(\frac{\theta}{2\pi}\right)^{\beta-k}\left(\frac{\rho}{\pi}\right)^{k}[\sin(\theta-\mu)+\sin(\mu)]^{k}.
\end{equation*}
						
Further, the  term $[\sin(\theta-\mu)+\sin(\mu)]^{k}$ may be rewritten as

\begin{align*}
&[\sin(\theta-\mu)]^{k}\left[1+\frac{\sin(\mu)}{\sin(\theta-\mu)}\right]^{k}M_0+[\sin(\mu)]^{k}\left[1+\frac{\sin(\theta-\mu)}{\sin(\mu)}\right]^{k}M_1\\ \\
=&[\sin(\theta-\mu)]^{k}\sum_{s=0}^{+\infty}{k \choose s}\left[\frac{\sin(\mu)}{\sin(\theta-\mu)}\right]^{s}M_0+[\sin(\mu)]^{k}\sum_{s=0}^{+\infty}{k \choose s}\left[\frac{\sin(\theta-\mu)}{\sin(\mu)}\right]^{s}M_1,
\end{align*}
where,  $M_0=I(|\sin(\theta-\mu)|\geq|\sin(\mu)|)$, $M_1=I(|\sin(\theta-\mu)|<|\sin(\mu)|)$, $I(.)$ refers to the indicator function and the portion composed by the quotients are, in module, less than $1$. Thus, from the Binomial theorem, it follows that
\begin{equation*}
F(\theta)=\sum_{k=0}^{+\infty}\sum_{s=0}^{k}{\beta \choose k}{k \choose s}\left(\frac{\theta}{2\pi}\right)^{\beta-k}\left(\frac{\rho}{\pi}\right)^{k}\left\{\frac{\sin(\mu)^s}{\sin(\theta-\mu)^{s-k}}M_0+\frac{\sin(\theta-\mu)^s}{\sin(\mu)^{s-k}}M_1\right\}.
\end{equation*}
						
To simplify this expression, we adopt the notation 
			
\begin{equation*}
T_{k,s}(\beta,\rho,\mu)={\beta \choose k}{k \choose s}\left(\frac{1}{2\pi}\right)^{\beta-k}\left(\frac{\rho}{\pi}\right)^{k}[\sin(\mu)]^{s}.
\end{equation*}
Thus,
\[
F(\theta)=\sum_{k=0}^{\infty}\sum_{s=0}^{k}T_{k,s}\theta^{\beta-k}[\sin(\theta-\mu)]^{s}\left\{[\sin(\theta-\mu)]^{k-2s}M_0+{[\sin(\mu)]^{k-2s}}M_1\right\}.
\]
Now, using integration by parts, the EC $p$th central trigonometric moment is given by 
\begin{align*}
\mu_{p}=&\mathbb{E}\{\cos[p(\Theta-\mu)]\}+i\mathbb{E}\{\sin[p(\Theta-\mu)]\}\\
=&\int_{0}^{2\pi} \cos[p(\theta-\mu)] dF(\theta)+i\int_{0}^{2\pi} \sin[p(\theta-\mu)] dF(\theta)\\
=&\cos(p\mu)+\int_{0}^{2\pi} p \{[\sin[p(\theta-\mu)]\} F(\theta) d\theta-i\left\{\sin(p\mu)+\int_{0}^{2\pi} p \{[\cos[p(\theta-\mu)]\} F(\theta) d\theta\right\}.
\end{align*}
						
The first moment is given by
\begin{equation*}
\mu_{1}=\mathbb{E}[\cos(\Theta-\mu)]+i\mathbb{E}[\sin(\Theta-\mu)],
\end{equation*}
where
\begin{equation*}
\mathbb{E}[\cos(\Theta-\mu)]=\cos(\mu)+  \sum_{k=0}^{\infty}\sum_{s=0}^{k}T_{k,s}\left\{A(\beta-k,0,k-s+1)M_0+{[\sin(\mu)]^{k-2s}}A(\beta-k,0,s+1)M_1\right\},
\end{equation*}
\begin{equation*}
\mathbb{E}[\sin(\Theta-\mu)]=-\sin(\mu)- \sum_{k=0}^{\infty}\sum_{s=0}^{k}T_{k,s}\left\{A(\beta-k,1,k-s)M_0-{[\sin(\mu)]^{k-2s}}A(\beta-k,1,s)M_1\right\}
\end{equation*}
and
\begin{equation*}
A(a,b,c)=\int_{0}^{2\pi}\theta^{a}[\cos(\theta-\mu)]^{b}[\sin(\theta-\mu)]^{c} d\theta.
\end{equation*}
						
\section{Dataset}
\begin{table}[h]
\centering
\caption{$21$ wind directions, at a weather station in Milwaukee, at 6.00am, on each of consecutive days [\cite{Data}].}
{\begin{tabular}{ccccccccccccc} 
\midrule 
$356$ & $97$  & $211$ & $232$ & $343$ & $292$ & $157$ & $302$ & $335$ & $302$ & $324$ \\  $85$ &
$324$ & $340$ &  
$157$ & $238$ & $254$ & $146$ & $232$ & $122$ & $329$  &  
\\  
\midrule
\end{tabular}}
\label{sample-table}
\end{table}


\begin{thebibliography}{}
								
								\bibitem[\protect\citeauthoryear{Abe and Pewsey}{2013}]{Abe}
								Abe, T. and Pewsey, A. (2013).
								Extending Circular Distributions through Transformation of Argument. \textit{Annals of the Institute of Statistical Mathematics}
								\textbf{65}, 833--858. 
								
								\bibitem[\protect\citeauthoryear{Abe and Pewsey}{2009}]{Sine}
								Abe, T. and Pewsey, A. (2009).
								Sine-skewed Circular Distributions. \textit{Statistical Papers}
								\textbf{52}, 683--707. 
								
								\bibitem[\protect\citeauthoryear{Abe et. al}{2009}]{Papa}
								Abe, T., Pewsey, A. and Shimizu, K. (2009).
								On Papakonstantinou's extension of the cardioid distribution. \textit{Statistics and Probability Letters}
								\textbf{79}, 2138--2147. 
								
								\bibitem[\protect\citeauthoryear{AL-Hussaini and Ahsanullah}{2015}]{Exponentiated}
								AL-Hussaini E. K. and Ahsanullah M. (2015). \textit{Exponentiated Distributions}. Paris: Atlantis Press.
								
								
								\bibitem[\protect\citeauthoryear{Batschelet}{1981}]{Bat}
								Batschelet, E. (1981). \textit{Circular Statistics in Biology}. New York: Academic Press.
								
								
								\bibitem[\protect\citeauthoryear{Boles and Lohmann}{2003}]{Zoo}
								Boles, L.C. and Lohmann, K. J. (2003).
								True Navigation and Magnetic Maps in Spiny Lobsters. \textit{Nature}
								\textbf{421}, 60--63.
								
								\bibitem[\protect\citeauthoryear{Box and Pierce}{1970}]{Independence}
								Box, G. E. P. and Pierce, D. A. (1970).
								Distribution of Residual Autocorrelations in Autoregressive-Integrated Moving Average Time Series Models. \textit{Journal of the American statistical Association}
								\textbf{65}, 1509--1526. 
								
								\bibitem[\protect\citeauthoryear{
									Fern\'andez-Dur\'an}{2007}]{Duran}
								Fern\'andez-Dur\'an J. J (2007).
								Circular Distributions Based on Nonnegative Trigonometric Sums. \textit{Biometrics}
								\textbf{60}, 499--503. 
								
								\bibitem[\protect\citeauthoryear{Fisher}{1993}]{Fisher}
								Fisher, N. I. (1993). \textit{Statistical analysis of circular data}. Cambridge: Cambridge University Press.
							
								
								
								\bibitem[\protect\citeauthoryear{
									Gatto and Jammalamadaka}{2007}]{Gatto}
								Gatto R. and Jammalamadaka S. R. (2007).
								The Generalized von Mises Distribution. \textit{Statistical Methodology}
								\textbf{4}, 341--353. 
								
								\bibitem[\protect\citeauthoryear{Jammalamadaka and SenGupta}{2001}]{Jama}
								Jammalamadaka, S. R. and SenGupta, A. (2001). \textit{Topics in Circular Statistics}. Singapore: World Scientific.
								
								
								\bibitem[\protect\citeauthoryear{Jammalamadaka and Sengupta}{1972}]{Geo}
								Jammalamadaka, S.R. and Sengupta, A. (1972).
								Mathematical Techniques for Paleocurrent Analysis: treatment of Directional Data. \textit{Journal of the International Association for Mathematical Geology}
								\textbf{4}, 235--248.
								
								
								\bibitem[\protect\citeauthoryear{Jeffreys}{1961}]{Jeffreys}
								Jeffreys, H. (1961). \textit{Theory of Probability}, 3nd edn. Oxford: Clarendon Press.
								
								
								\bibitem[\protect\citeauthoryear{Johnson and Wehrly}{1977}]{Data}
								Johnson, R. A. and Wehrly, T. E. (1977).
								Measures and models for angular correlation and angular-linear correlation. \textit{Journal of the Royal Statistical Society}
								\textbf{39}, 222--229. 
								
								\bibitem[\protect\citeauthoryear{Jones and Pewsey}{2012}]{Jones}
								Jones, M. C. and Pewsey, A. (2012).
								Inverse Batschelet Distributions for Circular Data. \textit{Biometrics}
								\textbf{68}, 183--193. 
								
								\bibitem[\protect\citeauthoryear{Kato and Jones}{2010}]{Kato}
								Kato, S. and Jones, M. C. (2010).
								A family of distributions on the circle with links to, and applications
								arising from, M\"obius transformation. \textit{Journal of the American Statistical Association}
								\textbf{105}, 249--262. 
								
								\bibitem[\protect\citeauthoryear{Mardia}{1972}]{Mardia}
								Mardia, K. V. (1972). \textit{Statistics of Directional Data}. London-New York: Academic Press.
								
								
								\bibitem[\protect\citeauthoryear{Pewsey et. al}{2013}]{Pewseybook}
								Pewsey, A., Neuh{\"a}user, M. and Ruxton, G. D. (2013). \textit{Circular statistics in R}, 3nd edn. Oxford: Oxford University Press.
								
								
								\bibitem[\protect\citeauthoryear{R Core Team}{2015}]{R}
								R Core Team (2015). \textit{R: A Language and Environment for Statistical Computing}. Vienna, Austria.
								
							\end{thebibliography}
\end{document}